\documentclass{emulateapj}
\usepackage{amsmath}
\usepackage{graphicx}
\usepackage{natbib}
\usepackage[scaled]{helvet}
\usepackage{epsfig}
\usepackage{url}
\bibpunct{(}{)}{;}{a}{}{,}
\usepackage{textcomp}
\usepackage{mathpazo}
\usepackage{xspace}

\usepackage{color}
\usepackage[normalem]{ulem}

\newcommand{\msun}{\ensuremath{\,M_\Sun}\xspace}
\newcommand{\rsun}{\ensuremath{\,R_\Sun}\xspace}

\newcommand{\kms}{{\,km\,s$^{-1}$}\xspace}

\newcommand{\minus}{\scalebox{0.75}[1.0]{$-$}}
\newcommand{\thisstar}{RW~Aur\xspace}

\begin{document}

\title{Multiple Stellar Fly-Bys Sculpting the Circumstellar Architecture in RW Aurigae}
\author{Joseph E. Rodriguez$^1$, 
Ryan Loomis$^1$, 
Sylvie Cabrit$^{2,3}$,
Thomas J. Haworth$^4$,
Stefano Facchini$^5$,
Catherine Dougados$^{3}$,
Richard A. Booth$^6$,
Eric L. N. Jensen$^{7}$,
Cathie J. Clarke$^6$,
Keivan G. Stassun$^{8,9}$,
William R. F. Dent$^{10}$,
J\'{e}r\^{o}me Pety$^{11,2}$\\
}

\affil{$^{1}$Harvard-Smithsonian Center for Astrophysics, 60 Garden St, Cambridge, MA 02138, USA}
\affil{$^{2}$Sorbonne Universit\'e, Observatoire de Paris, Universit\'e PSL, CNRS, LERMA, F75014 Paris, France}
\affil{$^{3}$Univ. Grenoble Alpes, CNRS, IPAG, 38000 Grenoble, France}
\affil{$^{4}$Astrophysics Group, Imperial College London, Blackett Laboratory, Prince Consort Road, London SW7 2AZ, UK}
\affil{$^{5}$Max-Planck-Institut f\"ur Extraterrestrische Physik, Giessenbachstrasse 1, D-85748 Garching, Germany}
\affil{$^{6}$Institute of Astronomy, University of Cambridge, Madingley Road, Cambridge, CB3 0HA, UK}
\affil{$^{7}$Department of Physics and Astronomy, Swarthmore College, Swarthmore, PA 19081, USA}
\affil{$^{8}$Department of Physics and Astronomy, Vanderbilt University, Nashville, TN 37235, USA}
\affil{$^{9}$Department of Physics, Fisk University, Nashville, TN 37208, USA}
\affil{$^{10}$Joint ALMA Observatory, Alonso de Córdova 3107, 763-0355 Vitacura, Santiago, Chile}
\affil{$^{11}$I.R.A.M., 300 rue de la Piscine, Domaine Universitaire, F-38406 Saint Martin d’Hres, France}
\shorttitle{RW Aurigae ALMA}
\shortauthors{Rodriguez et al.}

\begin{abstract}
We present high-resolution ALMA Band 6 and 7 observations of the tidally disrupted protoplanetary disks of the RW Aurigae binary. 
Our observations reveal the presence of additional tidal streams to the previously observed tidal arm around RW Aur A. The observed configuration of tidal streams surrounding RW Aur A and B is incompatible with a single star--disk tidal encounter, suggesting that the RW Aurigae system has undergone multiple fly-by interactions. We also resolve the circumstellar disks around RW Aur A and B, with CO radii of 58 au and 38 au consistent with tidal truncation, and 2.5 times smaller dust emission radii. The disks appear misaligned by 12\degr or 57\degr. Using new photometric observations from the American Association of Variable Star Observers (AAVSO) and All Sky Automated Survey for SuperNovae (ASAS-SN) archives, we have also identified an additional dimming event of the primary that began in late 2017 and is currently ongoing. With over a century of photometric observations, we are beginning to explore the same spatial scales as ALMA.

\end{abstract}


\section{Introduction}
The evolution of the circumstellar environment of a T Tauri star (TTS) from gas and dust to planets involves complex dynamical processes that are directly influenced by the presence of companions. It is known that most TTSs are in binaries \citep{Ghez:1993, Leinert:1993, Richichi:1994, Simon:1995, Ghez:1997}, and the process of planet formation can be altered and disrupted due to a stellar companion gravitationally influencing the gas and dust within the star's circumstellar disk. Specifically, strong binary interactions will stir up the disk, enhancing planetesimal collisions.  Additionally, binary interactions can excite the orbital eccentricities and inclinations of planetesimals as shown by stellar fly-by models to explain the outer structure of the the Kuiper belt in our own solar system \citep{Ida:2000}. Theoretical models predict that young binary systems may truncate the surrounding circumstellar material at a distance of up to 3 times the orbital separation of the two stars \citep{Artymowicz:1994}. This might explain the deficiency of debris disks around main sequence binaries with separation between 3--50 AU when compared to other single star systems and other binaries \citep{Trilling:2007}.

\begin{figure*}[!ht]
\vspace{0.3in}
\centering\includegraphics[width=0.99\linewidth, trim = 0 4.0in 0 0]{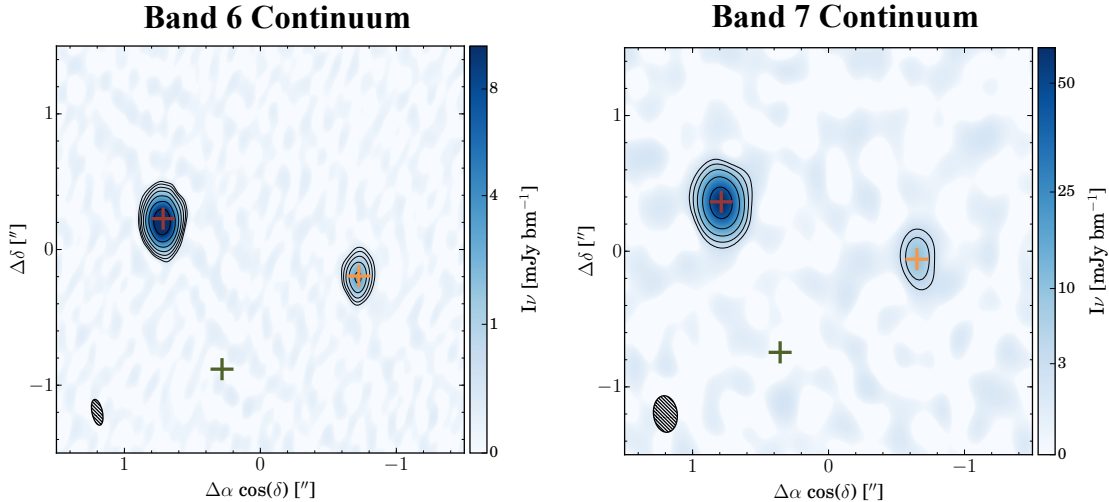}
\caption{Synthesized images of the Band 6 and 7 mm dust continuum emission. Contours in both panels correspond to [5,10,20,40,80,160,320,640]$\times\sigma$. The synthesized beam is 0$\farcs$19$\times$0$\farcs$08 in the Band 6 image and 0$\farcs$27$\times$0$\farcs$18 in the Band 7 image. The red, orange, and green crosses correspond to the center of the $^{12}$CO peaks for RW Aur A, B, and a third emission source, respectively.}
\label{figure:Continuum}
\end{figure*}

A demonstration of the impact of binary interactions on disk evolution (and possibly planet formation) is the classical TTS, RW Aurigae. The RW Aur system is 140 pc away \citep{vanLeeuwen:2007} and comprised of at least two stellar objects, RW Aur A and B, separated by $\sim$1.5$\arcsec$ \citep{Herbig:1988, Duchene:1999, StoutBatalha:2000}. RW Aur A has a large bipolar jet extending out to $\sim$100$\arcsec$ and containing many emission knots \citep{Mundt:1998, LopezMartin:2003}. Using the Plateau de Bure Interferometer (PdBI), this system was mapped in $^{12}$CO and dust continuum by \citet{Cabrit:2006} down to a resolution of 0.9$\arcsec\times0.6\arcsec$. Several unusual features were detected in these observations: a very compact rotating disk around RW Aur A, a large $\sim$600 AU long $^{12}$CO tidal arm stretching from it, and a circumstellar structure with complex kinematics around RW Aur B. From comparison with early numerical simulations by \citet{Clarke:1993}, \citet{Cabrit:2006} proposed that RW Aur B recently had a close encounter with RW Aur A, tidally stripping the original circumstellar disk around A. Although the disk around RW Aur A was not spatially resolved, comparison of its CO profile with theoretical models for Keplerian disks \citep{Beckwith:1993} indicated a small radius of 40-57 AU, inclined  45$^{\circ}$ to 60$^{\circ}$ to the line of sight. It was also speculated that the eccentric fly-by of RW Aur B possibly contributed to a temporary enhancement of the disk accretion rate onto RW Aur A ($4 \times 10^{-8}$ - $2 \times 10^{-7} M_\odot$ yr$^{-1}$, \citealp{Hartigan:1995, Facchini:2016}). Whether the A disk will survive this enhanced accretion episode is unclear: at the current rate, and with the disk mass estimated from dust ($3 \times 10^{-4} M_\odot$ \citealp{Andrews:2005}), the disk would be accreted in only 1000 yrs. 

\begin{figure*}[!ht]
\vspace{0.3in}
\centering\includegraphics[width=0.99\linewidth, trim = 0 0in 0 0]{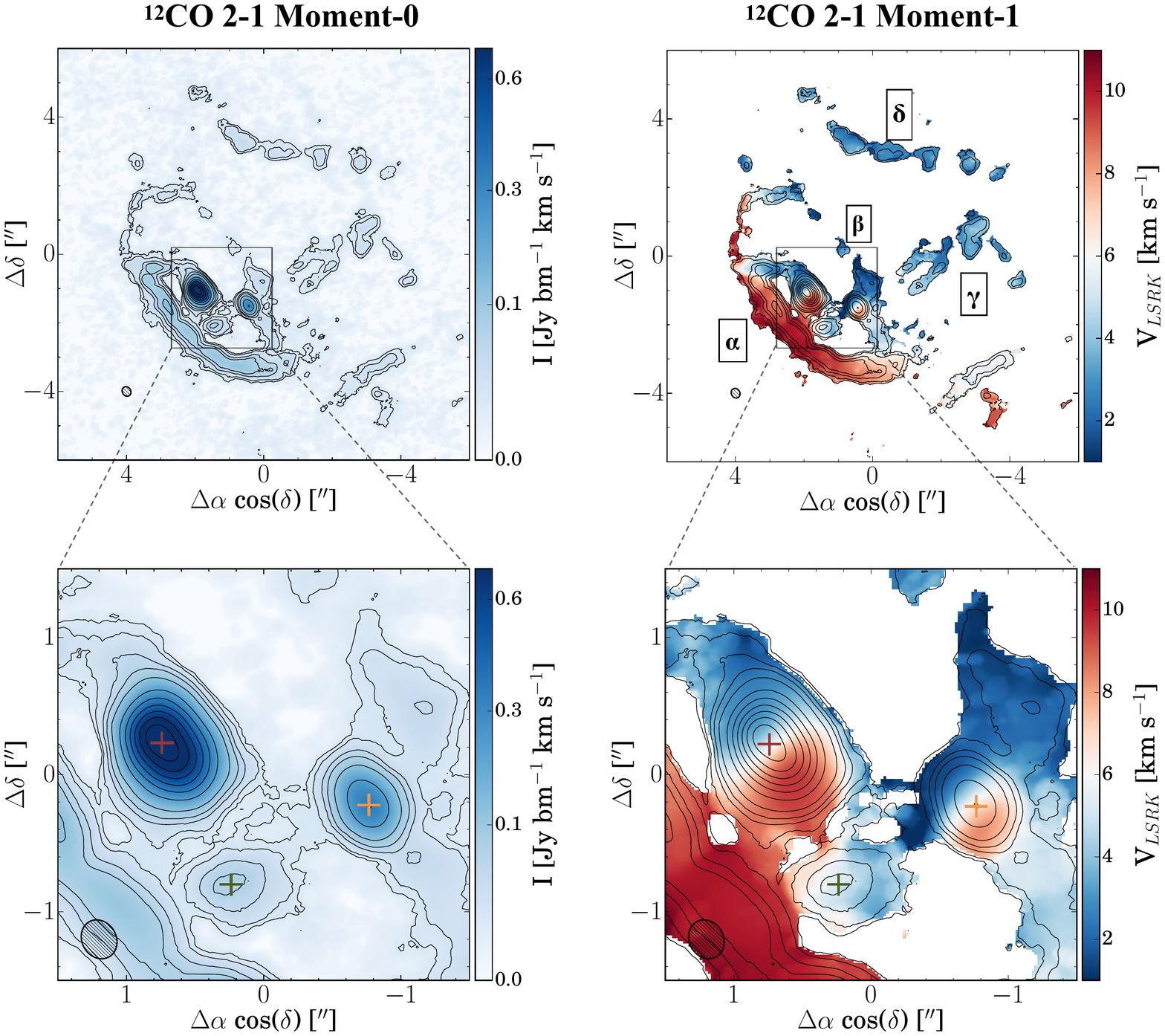}
\caption{Imaged $^{12}$CO(2-1) observations. (Left) Moment-0 map, with emission integrated from -1 to 13 \kms. A zoom-in of the center 3$\arcsec\times3\arcsec$ is shown below. Contours correspond to [3,5,10,15,25,50,75,100,125,150,175,200]$\times\sigma$. The synthesized beam is 0$\farcs$30$\times$0$\farcs$25. (Right) Moment-1 map overlaid with moment-0 contours, and a zoom-in of the center 3$\arcsec\times3\arcsec$ is shown below. The Kinematic Local Standard of Rest (LSRK) velocity of RW Aur A is 6 \kms. The original tidal arm discovered by \citet{Cabrit:2006} is labeled ``$\alpha$", the counter spiral arm is labeled ``$\delta$" and the two new tidal streams identified are labeled ``$\delta$" and ``$\gamma$". The red, orange, and green crosses in the bottom panels correspond to the center of the $^{12}$CO peaks for RW Aur A, B, and the third emission source, respectively.}
\label{figure:Band6CO}
\end{figure*}


\begin{figure*}[!ht]
\vspace{0.3in}
\centering\includegraphics[width=0.99\linewidth, trim = 0 4.3in 0 0]{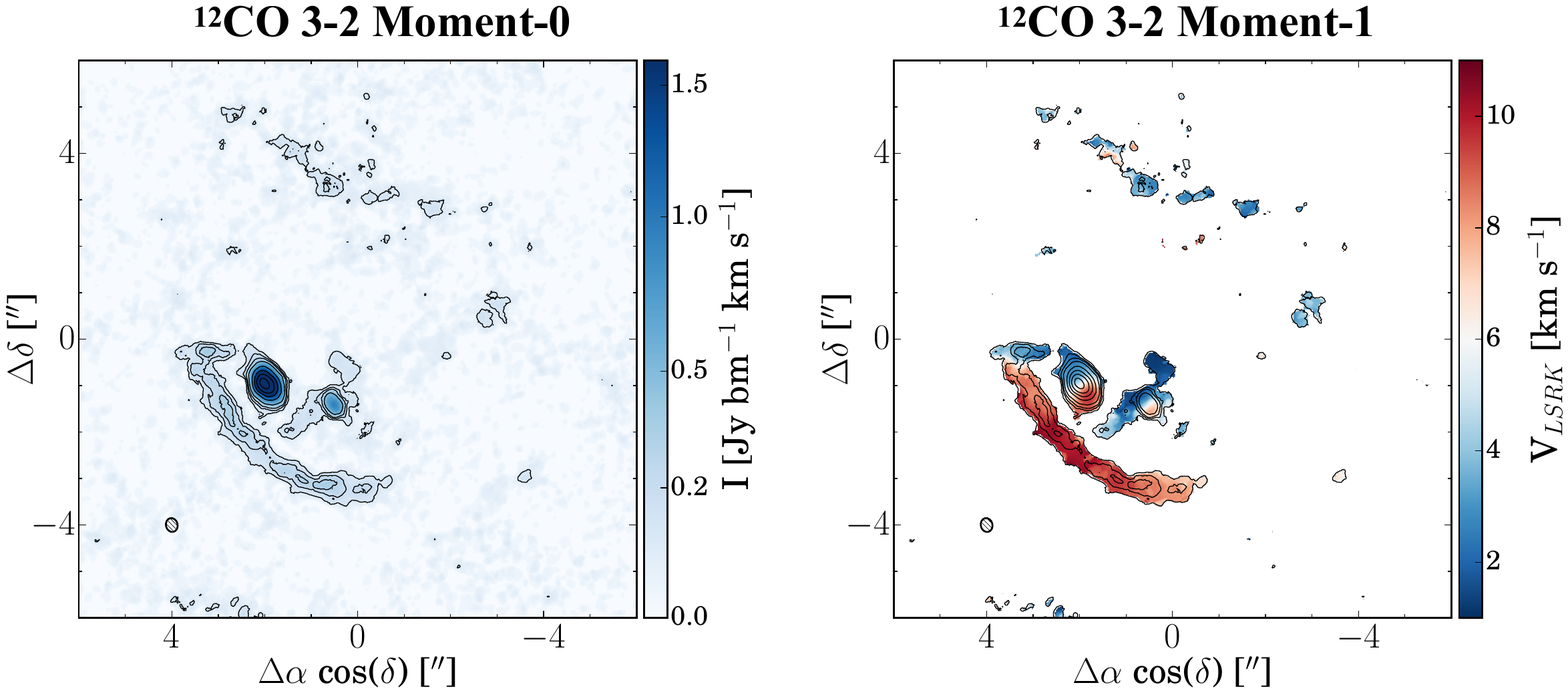}
\caption{Imaged $^{12}$CO(3-2) observations. (Left) Moment-0 map, with emission integrated from -1 to 13 \kms. Contours correspond to [3,5,10,15,25,50,75,100,125,150,175,200]$\times\sigma$. The synthesized beam is 0$\farcs$30$\times$0$\farcs$25. (Right) Moment-1 map overlaid with moment-0 contours. The systemic velocity of RW Aur A is 6 \kms.}
\label{figure:Band7CO}
\end{figure*}

To add to the mystery, RW Aur has also been exhibiting strong occultation events. After being photometrically monitored for over a century \citep{Beck:2001}, the RW Aur system faded by $\sim$2 mag in late 2010 for 180 days \citep{Rodriguez:2013}. From analyzing over 110 years of B-band photometric observations (photoelectric, photographic, and visual observations), no event similar in duration and depth was seen but the nominal brightness of RW Aur does vary on decade timescales \citep{Berdnikov:2017}.  Using basic kinematics, it was determined that the occulting object was moving at a few \kms and was likely $\sim$0.3 AU in diameter. Three years later, another large dimming ($>$4.5 mag) occurred, lasting $\sim$2 years \citep{Petrov:2015, Rodriguez:2016A, Lamzin:2017}. Near-IR observations of the RW Aurigae system suggest the occulting body consists of large grains ($\ge$ 1 $\mu$m), causing gray absorption \citep{Schneider:2015}. During the more recent dimming, there was an observed excess in the Near IR ($L$ and $M$) attributed to hot dust ($\sim$1000 K) in the inner disk \citep{Shenavrin:2015}. VRI polarimetry observations during the 2014-2016 dimming show a 20-30\% increase in polarization, consistent with what has been seen for UX Ori stars \citep{Lamzin:2017}. The high accretion rate of RW Aur A remained constant during both dimming events \citep{Chou:2013, Shenavrin:2015}. The cause of these unexpected dimming events is likely an occultation by a dust screen, but the origin of this screen is unclear and debated \citep{Rodriguez:2013, Petrov:2015, Shenavrin:2015, Bozhinova:2016, Facchini:2016, Berdnikov:2017}.

To further investigate the possibility that the unusual morphology of RW Aur might be explained by a tidal encounter in the form of a star-disk fly-by, \citet{Dai:2015} used a series of hydrodynamic (SPH) simulations to identify the orbital parameters that best reproduce all the features of RW Aur inferred from the PdBI observations. This included the length of the tidal arm, separation between RW Aur A and B, disk position angles, and relative stellar proper motions found by \citet{Bisikalo:2012}. Furthermore, they post-processed the hydrodynamic simulations with the non-LTE radiative transfer code \textsc{torus} (\citealp[e.g.][]{Harries:2000, Rundle:2010}) to compute synthetic molecular line and dust continuum observations to compare with the observations from \citet{Cabrit:2006}, finding excellent agreement with the CO optical depths, kinematic signatures in the line profiles, and observed continuum and CO flux densities, although CO emission around B is slightly underestimated.  The model also predicts that the line of sight to RW Aur A currently intersects a bridge of stripped off material between the two stars. Although of low column density ($N_H \le 10^{-4}$\,g cm$^{-2}$, i.e.\ $A_V \le 0.1$ mag), it was argued that the bridge structure may have small clumps of denser material able to occasionally dim the central star \citep{Dai:2015}.

In this paper, we present new Atacama Large Millimeter/submillimeter Array (ALMA) observations of the RW Aurigae system showing the aftermath of the star--disk interaction at high angular resolution. The new observations show the presence of additional tidal streams, suggesting the occurrence of multiple fly-bys of RW Aur B. Using new photometric observations by the American Association of Variable Star Observers (AAVSO) and the All Sky Automated Survey for SuperNovae (ASAS-SN), we analyze the full 2014-2016, 2016-2017, and the currently ongoing 2017-2018 dimming events. We also present an analysis of the archival AAVSO observations prior to 1960, showing a large dimming event that occurred in the late 1930s. Lastly, we announce that after a short period of quiescence after the 2016-2017 dimming, RW Aur began to fade again in mid-November of 2017. These new observations help to shed light on the intriguing nature of RW Aur itself, as well as the more general links between disk evolution and planet formation in binary systems. 

The paper is organized in the following way: Our ALMA and photometric observations are presented in \S\ref{Obs}. Our analysis of the high-resolution ALMA data is discussed in \S\ref{ALMA}. We discuss all six observed dimming events and the derived kinematics in \S\ref{Phot}. In \S\ref{discussion} we explore the impact these new results have on our understanding of the RW Aurigae system. Our results and conclusions are summarized in \S\ref{conclusion}. 

\begin{table*}
 \centering
 \caption{Results of Continuum and Spectral Line Fitting}
\small
\label{tbl:Results}
 \begin{tabular}{lcc}
    \hline
    \hline
    &RW Aur A & RW Aur B \\
    \hline
Band 6 Continuum 225 GHz & Beam PA = -12$^{\circ}$ & Beam Size = 0$\farcs$194$\times$0$\farcs$077  \\
    \hline
Disk Position Angle (degrees)& 41.05$\pm$0.16  & 40.06$\pm$1.39\\
Disk Inclination (degrees)& 55.51$\pm$0.13  & 67.61$\pm$1.93\\
Disk Flux Density (mJy) & 36.31 $\pm$ 0.02  & 4.61$\pm$0.02 \\
Disk Gaussian radius (AU)& 21.28 $\pm$ 0.03  & 15.51 $\pm$ 0.20  \\
RA (J2000)& 05:07:49.57220$\pm$0.00034s &  05:07:49.46089$\pm$0.00034s  \\
DEC (J2000)& +30:24:04.9362$\pm$0.0095$\arcsec$ & +30:24:05.3454$\pm$0.0095$\arcsec$ \\
A--B Separation & 1.497$\pm$0.001$\arcsec$&---\\
A--B Position Angle (degrees) &254.14$\pm$0.02&---\\
\hline
Band 7 Continuum 338 GHz & Beam PA = -7$^{\circ}$& Beam Size = 0$\farcs$271$\times$0$\farcs$175 \\
\hline
Disk Position Angle (degrees)& 37.86$\pm$0.88  & 30.13$\pm$4.71\\
Disk Inclination (degrees)& 57.68$\pm$0.86  & 72.08$\pm$7.98 \\
Disk Flux Density (mJy)& 85.62 $\pm$ 0.31  & 12.18$\pm$0.28 \\
Disk Gaussian Radius (AU)& 23.06 $\pm$ 0.20  & 21.56 $\pm$ 1.48  \\
RA (J2000)& 05:07:49.57048$\pm$0.00069s &  05:07:49.45974$\pm$0.00070s  \\
DEC (J2000)& +30:24:04.9970$\pm$0.013$\arcsec$ & +30:24:05.380$\pm$0.014$\arcsec$ \\
A--B Separation & 1.490$\pm$0.010$\arcsec$&---\\
A--B Position Angle (degrees) &254.05$\pm$0.12&---\\
\hline
12CO(2-1) &  Beam PA = -30$^{\circ}$ & Beam Size = 0$\farcs$30$\times$0$\farcs$25 \\
\hline
Disk Radius & 58 AU & 38 AU \\
Int. Disk Flux Density (Jy \kms)& 3.42$\pm$0.03 & 0.86$\pm$0.03 \\
Major Axis Ring& 6.1$\arcsec$&---\\
Minor Axis Ring& 2.8$\arcsec$&---\\
B--C Separation & 1.2$\arcsec$&---\\
A--C Separation & 1.1$\arcsec$&---\\
\hline
12CO(3-2) Int. Disk Flux Density (Jy \kms) & 10.4$\pm$0.3 & 2.4$\pm$0.3 \\
13CO(2-1) Int. Disk Flux Density (Jy \kms) & 0.26$\pm$0.03 & 0.1$\pm$0.03 \\
13CO(3-2) Int. Disk Flux Density (Jy \kms) & 1.1$\pm$0.3 &0.35$\pm$0.3 \\
C18O(2-1) Int. Disk Flux Density (Jy \kms) &0.09$\pm$0.03 & $<$0.06 (2$\sigma$) \\
SO(2-1) Int. Disk Flux Density (Jy \kms) &$<$0.06 (2$\sigma$) & $<$0.06 (2$\sigma$) \\
\hline
 \end{tabular}
\begin{flushleft} 
 \footnotesize{ \textbf{\textsc{NOTES:} Measurements on the continuum were made by fitting two elliptical Gaussian disks in the (u,v) plane (see \S \ref{Obs}). Source C was not detected in the continuum. Errors on the gas flux densities are likely underestimated as distinguishing which feature is contributing flux is not trivial. }
}
\end{flushleft}
\end{table*} 






\begin{figure}[!ht]
\vspace{0.3in}
\centering\includegraphics[width=0.99\linewidth, trim = 0 0 0 0]{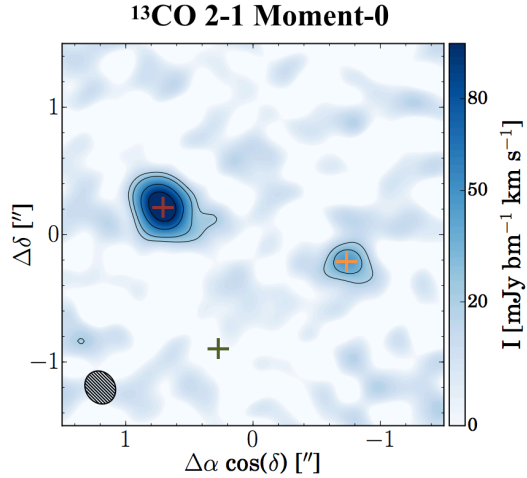}
\caption{Moment-0 map of $^{13}$CO(2-1) emission, integrated from -1 to 13 \kms. Contours correspond to [3,5,10,15]$\times\sigma$. The synthesized beam is 0$\farcs$27$\times$0$\farcs$23. The red, orange, and green crosses correspond to the center of the $^{12}$CO peaks for RW Aur A, B, and the third emission source, respectively.}
\label{figure:13CO}
\end{figure}

\section{Observations}
\label{Obs}
\subsection{ALMA}
As part of ALMA Cycle 3 and 4 projects 2015.1.01506.S and 2016.1.00877.S, RW Aur was observed in Band 6 (225 GHz) with the 12m array on UT 2016 September 29/30 and UT 2016 December 07/08 for a total integration time of 311.17 minutes. The observations included 39 antennas with a baseline of 15 -- 3248 m. The quasar J0510+1800 was used as the flux, phase, and bandpass calibrator for all scheduling blocks except one which used J0512+2927 as the phase calibrator. Also, Band 6 observations were taken with the 7m array for a total additional integration time of 217.23 minutes. This observation used 11 antennas with a baseline range of 9 -- 45 m. J0510+1800 was used as the bandpass and phase calibrators, and the quasar J0522-3627 was used for the flux calibrator.

Additionally, RW Aur was observed in Band 7 (338 GHz) on UT 2016 July 23 as part of the cycle 3 Alma proposal for a total integration time of 22.8 minutes with a 16 -- 1110 m baseline range. Only 36 antennas were used in this observation. J0510+1800 was used as both bandpass and flux calibrator while J0512+2927 was observed for phase calibration.

The data were initially calibrated by the NAASC. Two rounds of phase-only self-calibration were then applied to each set of observations in CASA 4.3.1 using the RW Aurigae continuum emission as the self-calibration model. The Band 6 observations were concatenated and imaged in CASA 4.3.1 using the CLEAN task. The continuum data were imaged using Briggs weighting with a robust value of 0, yielding a synthesized beam of 0$\farcs$194$\times$0$\farcs$077, a beam position angle of -12$^{\circ}$, and an rms of 38~$\mu$Jy bm$^{-1}$ (Figure \ref{figure:Continuum}). The $^{12}$CO 2--1 line was imaged at a velocity resolution of 0.5 \kms, a robust value of 0.5, and with a uv-taper of 0$\farcs$06$\times$0$\farcs$25 to force a more circular beam and improve signal to noise, resulting in a synthesized beam of 0$\farcs$30$\times$0$\farcs$25, a beam position angle of -30$^{\circ}$, and an rms of 1.5~mJy/bm (Figure \ref{figure:Band6CO}). The $^{13}$CO 2--1 line was identically imaged, resulting in a synthesized beam of 0$\farcs$27$\times$0$\farcs$23, a beam position angle of -32$^{\circ}$, and an rms of 2.0~mJy/beam per 0.5 km/s channel (Figure \ref{figure:13CO}).

The Band 7 continuum data were similarly imaged using Briggs weighting with a robust value of 0, yielding a synthesized beam of 0$\farcs$271$\times$0$\farcs$175, a beam position angle of -7$^{\circ}$, and an rms of 560~$\mu$Jy/bm (Figure \ref{figure:Continuum}). The $^{12}$CO 3--2 line was imaged at a velocity resolution of 0.5 \kms, a robust value of 0.5, and with a uv-taper of 0$\farcs$06$\times$0$\farcs$25 to match the 0$\farcs$30$\times$0$\farcs$25 beam from the Band 6 observations, yielding an rms of 13.5~mJy/bm and a beam position angle of -13$^{\circ}$ (Figure \ref{figure:Band7CO}). The $^{13}$CO 3--2 line was similarly imaged, resulting in a synthesized beam of 0$\farcs$32$\times$0$\farcs$26, a beam position angle of -15$^{\circ}$, and an rms of 17.5~mJy/beam per 0.5 \kms channel (not shown). Our final calibrated ALMA observations are available publicly with this paper.

\begin{figure}[!ht]
\vspace{0.3in}
\centering\includegraphics[width=\linewidth, trim = 0 0 0 0, angle = 0]{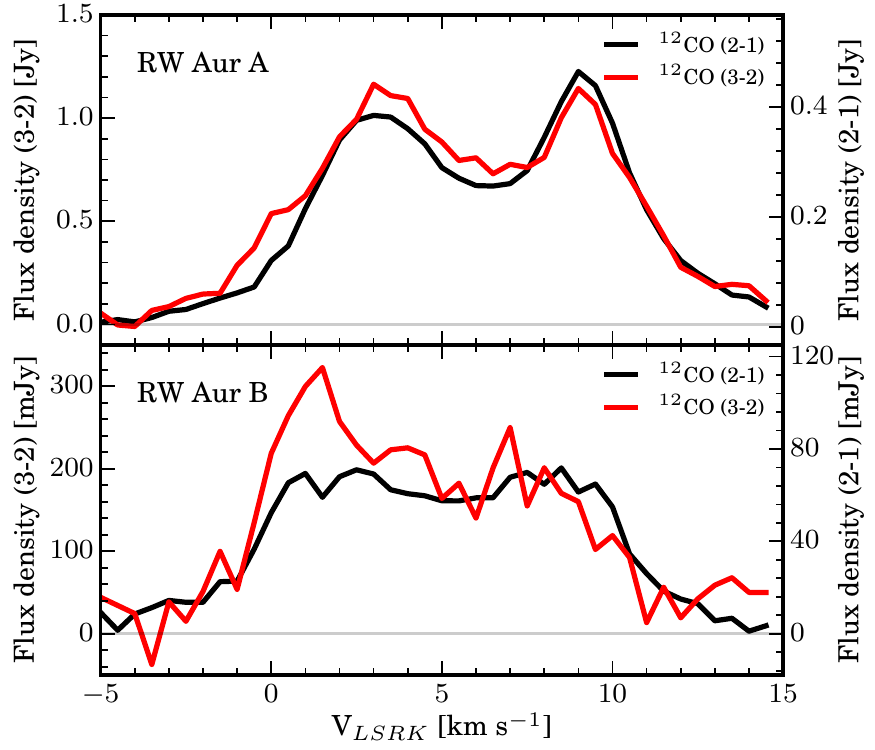}
\caption{$^{12}$CO spectra of RW Aur A (Top) and B (Bottom). The J = 2-1 spectrum is shown in black and the J = 3-2 spectrum is shown in red. The J = 2-1 spectra for RW Aur A and B have been multiplied by the (3-2)/(2-1) ratios in Table \ref{tbl:Results}. An elliptical mask with a size equivalent to the measured CO extent was used to extract the spectra.}
\label{figure:spectra}
\end{figure}

\subsection{Archival Observation: KELT \& Wesleyan University}
We include and analyze the archival observations of RW Aur from the Kilodegree Extremely Little Telescope (KELT) survey and the Wesleyan University's Van Vleck Observatory that were used to first identify that RW Aurigae was experience long duration, large amplitude dimming events \citep{Rodriguez:2013}. The Wesleyan photometric archive\footnote{\url{https://wesfiles.wesleyan.edu/home/wherbst/web/TTauriDataBase/}} for T-Tauri stars is described in \citet{Herbst:1994}. See \citet{Rodriguez:2013} for a description of the KELT observations on RW Aurigae.

\subsection{ASAS-SN}
Using two fully robotic units, each consisting of four telescopes, the All Sky Automated Survey for SuperNovae (ASAS-SN) is monitoring the entire sky down to a visual magnitude of $\sim$17 \citep{Shappee:2014, Kochanek:2017}. Each telescope is a 14cm Nikon telephoto lens equipped with a 2k $\times$ 2k thinned CCD and the 8 telescopes together can cover $\sim$20,000 deg$^{2}$ each night. The survey is designed to discover new supernovae and transient sources. Each telescope has a $8.8^{\circ}$ $\times$ $8.8^{\circ}$ field of view and a 7$\farcs$8 pixel scale. ASAS-SN observed RW Aur from UT 2015 March 03 to UT 2018 March 18, obtaining 591 observations in the $V$-band. The median per point error is 0.009 mag. 

\subsection{AAVSO}
Dedicated to the understanding of variable stars, the American Association of Variable Star Observers (AAVSO) is an amateur-professional network of CCD and visual observers worldwide. The AAVSO has photometric observations from UT 1906 December 18 to UT 2018 March 22, totaling 15,057 visual and V-band CCD observations on RW Aur. Only some of the observations have a reported uncertainty. The observations on RW Aurigae are publicly available for the community\footnote{https://www.aavso.org/}.

\section{ALMA Results}
\label{ALMA}
In this section, we describe the new cycle 3 and 4 ALMA observations of RW Aurigae that suggest we are observing the aftermath of multiple eccentric fly-by interactions by RW Aur B, first proposed by \citet{Cabrit:2006}. 


\begin{figure*}[!ht]
\vspace{0.3in}
\centering\includegraphics[width=0.99\linewidth, trim = 0 0in 0 0]{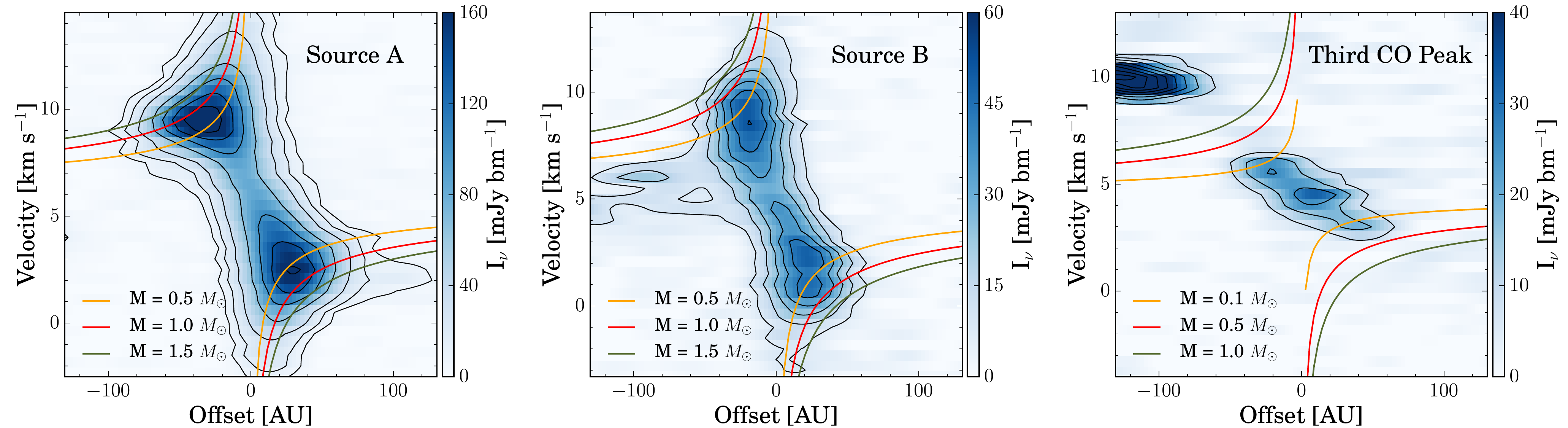}
\caption{Position-velocity diagrams of the $^{12}$CO(2-1) emission from RW Aur A (Left), B (Middle), and the third source we see between RW Aur A, B, and the tidal arm (Right). The data are extracted using position angles of 41$^{\circ}$ and  for RW Aur A and B, respectively, taken from our fit to the Band 6 continuum. A position angle of 270$^{\circ}$ was used for the third source, derived visually from the direction of apparent rotation. Contours correspond to [5,10,15,25,30,35,40]$\times\sigma$. Keplerian velocity profiles are shown for three different stellar masses for each source, using inclinations of 55$^{\circ}$, 67$^{\circ}$, and 52$^{\circ}$ and central velocities of 6, 5.2, and 4.5 \kms for RW Aur A, B, and the third source, respectively. }
\label{figure:PV}
\end{figure*}

\subsection{Continuum Modeling}
\label{continuum}
To trace the dust in the RW Aurigae system, we analyze both the Band 6 and 7 continuum observations (see Figure \ref{figure:Continuum}). These observations spatially resolve the disks around both RW Aur A and B, the first time RW Aur B has been resolved. To best interpret this emission, we fit a simple model consisting of two Gaussian disks directly to the measured visibilities. For each disk, we fit for a position angle, inclination, integrated flux density, central offset, and a Gaussian width as a proxy for disk radius. This model used a total of 12 parameters and was fit to the observed visibilities using the MCMC routine \texttt{emcee} \citep{ForemanMackey:2013} combined with the visibility sampling routine \texttt{vis\_sample} \citep{Loomis:2018}\footnote{\url{vis\_sample is publicly available at https://github.com/AstroChem/vis\_sample or in the Anaconda Cloud at https://anaconda.org/rloomis/vis\_sample.}}. The best fit model resulted in a reduced $\chi^{2}$ of 1.02, and the fit parameters are shown in Table \ref{tbl:Results}.


The dust disk emission radius is compact around both sources, $\simeq$ 20 AU, with a 30\% smaller radius for source B at longer wavelengths. The A and B disks have similar PAs (within 1 sigma), with the disk around RW Aur A perpendicular to its optical jet (PA = $130^{\circ} \pm 2^{\circ}$, \citealp{Dougados:2000}). In contrast, the disk inclinations differ by 12$^{\circ}$ (6 sigma) in Band 6 (and $15^{\circ}\pm 8^{\circ}$ in Band 7). It seems difficult to attribute this difference to residual random phase noise, as we would expect this effect to make the smaller disk (B) appear rounder and less inclined than the larger one (A). We therefore conclude that the disks around RW Aur A and B appear to be misaligned relative to each other, by roughly 12$^{\circ}$ if their upper surfaces both face to the SE (the direction of the blueshifted jet of RW Aur A), or by 57$^{\circ}$ if they face in opposite directions. Other young binary systems where both stars have a circumstellar disk have also been shown to be misaligned relative to each other \citep[e.g.,][and references therein]{Jensen:2014}.


Our disk fits give a separation between RW Aur A and B of 1.497$\pm$0.001$\arcsec$ in Band 6 and 1.490$\pm$0.010$\arcsec$ in Band 7, which is in agreement with the AB separation measured by \citet{Cabrit:2006} of 1.468$\pm$0.056$\arcsec$ but larger than the optical HST separation of 1.4175$\pm$0.0034$\arcsec$ \citep{Ghez:1997}. The separation of RW Aur A and B was also measured with Chandra to be 1.48$\pm$0.01$\arcsec$ \citep{Skinner:2014}. Recently, Gaia measured the separation to be 1.4863$\pm$0.0006$\arcsec$ \citep{Gaia:2016}. Using a large compilation of data from the literature, \citet{Csepany:2017} determined that RW Aur A and B are currently moving away from each other at about 3.6 mas yr$^{-1}$ (see their Tables 2 and 5); accounting for this increase in separation puts our observations in excellent agreement with the HST data of \citet{Ghez:1997}, and with the other accurate separation measurements obtained in the meantime with VLT/NACO \citep{Correia:2006}, CHANDRA \citep{Skinner:2014}, and Gaia \citep{Gaia:2016}.


Our total Band 6 flux density for the RW Aur A and B system is 40.92$\pm$0.03 mJy (RW Aur A = 36.31 $\pm$ 0.02 mJy and RW Aur B = 4.61$\pm$0.02 mJy), which is consistent with the 1.3mm total system flux density of 42$\pm$2 mJy measured by \citet{Osterloh:1996} but 25\% higher than the 32.8$\pm$0.7 mJy (RW Aur A = 27.6$\pm$0.5 mJy and RW Aur B = 5.2$\pm$0.4 mJy) determined by \citet{Cabrit:2006}. This discrepancy is compatible with the absolute flux calibration accuracy of $\simeq$ 20\% estimated in \citet{Cabrit:2006} for their PdBI observations. Their smaller flux ratio of A/B at 1.3mm (5, instead of 8 in the ALMA data) could result from the fact that RW Aur B was detected at a low signal to noise on top of a faint emission bridge connecting to RW Aur A, which may have caused an overestimate of the B flux density (this bridge is not seen in the more sensitive ALMA data, and was thus likely a cleaning artifact). 
In Band 7, our total measured flux density of 97.8$\pm$0.5 mJy is 20-30\% larger than the total system flux density at 863 microns of $70 \pm 4$ mJy measured with SCUBA in single-dish by \citet{Andrews:2005}, again consistent within the absolute flux calibration errors.


\subsubsection{Spectral Index}
Comparing the ALMA Band 6 (225 GHz) and Band 7 (338 GHz) flux densities, we find a spectral index ($\alpha = \Delta \log F_{\nu}$$/$$\Delta \log \nu$) of 2.11$\pm$0.01 in RW Aur A and 2.39$\pm$0.07 in RW Aur B. In both sources, the PdBI 122.5 GHz continuum flux densities from \citet{Cabrit:2006} are consistent with an extrapolation along these slopes, indicating a constant spectral index across this wavelength range for both disks. The shallower spectral index of 1.73$\pm$0.08 reported for RW Aur A by \citep{Cabrit:2006} was probably caused by their 25\% underestimate in absolute flux calibration at 1.3 mm (see discussion in previous paragraph). We note that for a disk of radius 20 AU viewed at a 55 degrees of inclination, the mean brightness temperature required to reproduce the observed continuum flux in RW Aur A in Band 7 is about 39 K; this value matches very well with the dust temperature at 20 AU predicted by fits to the infrared SED of RW Aur A, T$_d$(r) = 217 K (r / AU)$^{\minus0.57}$ \citep{Osterloh:1995}. Hence, the dust emission in RW Aur A may be optically thick at 338 GHz. This would be consistent with the spectral slope of 1.43$\pm$0.21 reported by \citet{Andrews:2005} for the total system flux density (dominated by A) between 850 $\mu$m and 380 $\mu$m (a blackbody at 39 K predicts an index of 1.6 in this frequency range). At longer wavelengths, the observed spectral index $\alpha \simeq 2.11$ between ALMA Band 7 and Band 6 is slightly steeper than the value of 1.8 predicted for a 39 K blackbody, indicating that the emission is becoming optically thin below 200 GHz\footnote{For example, if we crudely approximate the disk as a greybody at 39 K, the observed flux densities could be reproduced with an optical depth $\tau_\nu \simeq 3 (\nu/338 GHz)^{0.8}$}. In RW Aur B, the 8 times weaker flux density for a similar dust emission radius $\simeq$ 20 AU indicates that dust continuum is optically thin even in Band 7, consistent with the steeper slope of 2.4.


\subsection{Spectral Line Emission}
\label{SpectralLine}
To map out the kinematics and size distributions of the gas in the RW Aurigae system, we analyze the $^{12}$CO(2-1, 3-2), $^{13}$CO(2-1, 3-2), (see Figures \ref{figure:Band6CO}, \ref{figure:Band7CO}, \& \ref{figure:13CO}) C$^{18}$O(2-1) and SO(2-1) emission (not shown). Only a small amount of C$^{18}$O(2-1) emission is detected toward RW Aur A, and none is detected toward RW Aur B. We do not detect any SO(2-1) emission from either source. Our analysis of the $^{12}$CO(2-1) emission (see the Moment-1 map in Figure \ref{figure:Band6CO}) reveals the presence of numerous features: disks around RW Aur A and RW Aur B, the original tidal arm (labeled ``$\alpha$" in the moment-1 map), and a second blueshifted counter-arm forming an apparent ring of emission with the main one (labeled ``$\beta$" in the moment-1 map), a second tidal stream that is inclined relative to the apparent circumbinary ring (labeled ``$\delta$" in the moment-1 map), a third tidal stream (labeled ``$\gamma$" in the moment-1 map), and a third clump of emission originating between RW Aur A, B, and the original tidal arm. 

\subsubsection{RW Aur A}

RW Aur A is detected in $^{12}$CO (2-1 = 3.42$\pm$0.03 Jy \kms , 3-2 = 10.4$\pm$0.3 Jy \kms ), $^{13}$CO (2-1 = 0.26$\pm$0.03 Jy \kms , 3-2 = 1.1$\pm$0.3 Jy \kms ), and C$^{18}$O(2-1) (0.09$\pm$0.03 Jy \kms ). We place a 2$\sigma$ upper limit on SO (2-1) emission of $<$0.06 Jy \kms. 

A clear Keplerian velocity gradient is observed across the disk around RW Aur A, consistent with a stellar mass range of 1--1.5\msun, assuming a PA of 41$^{\circ}$ and a dust disk inclination of 55$^{\circ}$ from our (u,v) fit of the Band 6 continuum (see Figure \ref{figure:PV}). This mass range is consistent with the inferred mass of 1.3-1.4\msun determined from pre-main sequence models of near-IR observations \citep{Ghez:1997, Woitas:2001}. Using CASA's built-in measurement tool \citep{McMullin:2007}, we estimate the radius of the gas disk around RW Aur A from the $^{12}$CO(2-1) to be $\sim$58 AU, $\sim$2.5 times larger than the extent of the mm continuum emission.


$^{12}$CO (2-1) and (3-2) spectra (see Figure \ref{figure:spectra}) were extracted from the image cubes using an elliptical mask with a size equivalent to the measured CO extent of RW Aur A shown in Table \ref{tbl:Results}. The line profiles are double peaked, as expected for a Keplerian disk, but are not symmetric about the Kinematic Local Standard of Rest (LSRK) systemic velocity (estimated as $V_{LSRK}(A) \simeq 6$ \kms).  The blue-shifted side of the peak is broader and shallower than the red-shifted side relative to the motion of the star. This observed asymmetry in the integrated spectra may be caused by material being sheared off the disk into the tidal arm, as seen in the moment 1 maps. From our results, the ratio of $^{12}$CO (3-2)/(2-1) $\sim$ 3.0. 

The peak surface brightness in $^{12}$CO(2-1) of $S_\nu = 190$ mJy/beam translates into a peak brightness temperature $T_b = 58$ K. This indicates that the CO emitting gas at 60 AU in the RW Aur A disk is significantly warmer than the dust at 20 AU (for which continuum flux densities were found consistent with a temperature $\sim$ 40 K). This result confirms the conclusion of \citet{Cabrit:2006} that CO emission in the RW Aur A disk appears to probe a warm super-heated disk surface layer located above the cool midplane probed by mm dust emission. ALMA studies have clearly resolved the vertical increase in temperature in the atmosphere of disks on scales of 100-200 au \citep[][]{Pinte2018} but the tidal truncation of the disk in RW Aur allows us to probe this effect at much smaller radii. 



\subsubsection{RW Aur B}
RW Aur B was only detected in $^{12}$CO (2-1 = 0.86$\pm$0.03 Jy \kms , 3-2 = 2.4$\pm$0.3 Jy \kms ) and $^{13}$CO (2-1 = 0.1$\pm$0.03 Jy \kms , 3-2 = 0.35$\pm$0.3 Jy \kms ). The PV diagram for RW Aur B suggests Keplerian rotation corresponding to a stellar mass of 0.5\msun $<M<$ 1\msun, assuming an inclination of 79$^{\circ}$ and a PA of 46.4$^{\circ}$ from our continuum fitting (see Figure \ref{figure:PV}). This mass range is consistent with the inferred mass of 0.7--0.9\msun \citep{Ghez:1997, Woitas:2001}. We estimate the radius of the gas disk around RW Aur B in $^{12}$CO(2-1) to be 38 AU, $\sim$2.5 times larger than its dust disk. 

Although the integrated line profile is noisier for RW Aur B (see Figure \ref{figure:spectra}), we do detect a double peaked Keplerian profile for both $^{12}$CO J = 3-2 and J = 2-1 around a systemic velocity of about 5.2 \kms, blueshifted by -0.8 \kms\ from A. The observed $^{12}$CO (3-2)/(2-1) ratio is 2.8.


The peak surface brightness in $^{12}$CO(2-1) of $\sim 50$ mJy/beam translates into a brightness temperature $T_b =15$ K. The disk of RW Aur B is only marginally resolved by our beam (radius of 38 AU $\simeq 0.27\arcsec$). Therefore, the observed peak surface brightness probably suffers substantial beam dilution and the true brightness is higher than this.

\subsubsection{Tidal Streams}
\label{tidalstreams}
Observations from IRAM Plateau de Bure Interferometer discovered a long tidal stream wrapped around RW Aur A, suggesting that RW Aur had previously undergone an eccentric star--disk encounter \citep{Cabrit:2006}. These observations resembled the properties of simulated stellar fly-bys on an accretion disk \citep{Clarke:1993}. Therefore, it was proposed that RW Aur B is on a prograde, highly eccentric orbit causing it to come very close to RW Aur A, possibly within the extent of the original circumstellar disk around RW Aur A. It was determined that unlike previously observed circumbinary rings, the velocity structure of the arm was inconsistent with pure rotation. The NE and SW extension of the arm is blue-shifted while the middle of the arm is red-shifted. \citet{Cabrit:2006} suggested that the arm is likely expanding outward from RW Aur. 

The new $^{12}$CO emission maps (see Figure \ref{figure:Band6CO} and \ref{figure:Band7CO}) clearly show the presence of the original tidal arm (labeled ``$\alpha$" in the moment-1 map shown in Figure \ref{figure:Band6CO}) found by \citet{Cabrit:2006} and confirm that the NE extension is blue shifted while most of the arm is red shifted with respect to the systemic velocity. The ALMA observations resolve the arm, showing evidence of possible clumpy substructure along its length. Given the recent occultation events, which may suggest that material is coalescing, it is not surprising to see substructure within the disrupted gas shown in the $^{12}$CO maps. The arm is connected to the NE portion of the RW Aur A disk and is not detected in the continuum. From the $^{12}$CO(2-1) map we estimate the arm to be $\sim$740 AU along the spine of the arm ($\sim$5.2$\arcsec$), having an estimated typical width of $\sim$85 AU ($\sim$0.6$\arcsec$). We note that the arm fluctuates in its apparent width by $\sim$0.1$\arcsec$ and has a total integrated flux density in the $^{12}$CO(2-1) of 3661$\pm$58 mJy \kms. It appears that this tidal stream wraps fully around RW Aur A and B, forming the appearance of a circumbinary ring of gas. We note that this is likely a projection effect and that there is actually a second counter spiral around RW Aur B from its fly-by (labeled ``$\beta$"). In \S\ref{discussion} we discuss the nature and morphology of these tidal streams, providing evidence to favor the dual spiral arm interpretation.

In addition to the main tidal arm ($\alpha$) first identified by \citet{Cabrit:2006}, and the secondary arm $\beta$, we identify the presence of two additional tidal streams extending out past the extent of the apparent circumbinary ring (labeled ``$\gamma$" and ``$\delta$" in the moment-1 map shown in Figure \ref{figure:Band6CO}). For clarity, we refer to these two features as ``tidal streams'' in comparison to the two features that form the projected circumbinary ring, which we refer to as ``spirals'' or ``arms''. It is possible that these two new tidal streams are actually one continuous stream, but our observations are not sensitive enough to constrain this due to a lack of (u,v) coverage at intermediate baselines. In \S\ref{discussion} we discuss whether these additional tidal streams are bound to the system.

One of the new tidal streams ($\gamma$ in Fig. \ref{figure:Band6CO}) appears to be inclined relative to the tidal arm (possibly perpendicular). While this could be partly due to projection effects (foreshortening), we note that the orientation of the $\gamma$ stream on the sky is within a few degrees of perpendicular to the PA of the disk around RW Aur B. This coincidence raises the alternative possibility that $\gamma$ might trace a slow molecular outflow from RW Aur B. RW Aur B is also an active accretor \citep{Duchene:1999} and could have experienced an enhanced ejection phase near periastron passage. Additional (u,v) coverage at intermediate scales would be needed to test this scenario by recovering the full structure of this stream.

Finally, to the south of the two stars, we also note a bright clump of CO emission which we refer to as the ``third CO peak'' (see Figure \ref{figure:Band6CO}). The linear slope of the PV diagram for this emission clump (see Figure \ref{figure:PV}) is not consistent with resolved Keplerian rotation. On the other hand, it is reminiscent of the linear PV diagram across the (unresolved) B-disk in the PdBI observations of \citet{Cabrit:2006}. Also, a few PV diagrams with similar linear slopes are observed towards edge-on protostars, where they appear to trace a rotating ring near the disk centrifugal barrier (see e.g. Figure 5 in \citealp{Lee:2017}). The Keplerian rotation curves shown in Figure \ref{figure:PV} would then set an upper limit of $\sim$0.1 \msun\ to the mass of a possible embedded source in the third CO peak. 

To constrain the presence of continuum emission associated with the third $^{12}$CO peak, the Band 6 observations were re-imaged with natural weighting. No continuum emission was detected at this location, with a 2 sigma upper limit of 30 $\mu$Jy. Comparing this limit with the measured $^{12}$CO 2-1 integrated flux density (220 mJy km/s), we find a CO to continuum ratio of $>7000$ \kms, while RW Aur A/B have ratios of 94 and 187 \kms, respectively. 

We further note that the location of this emission would likely be very unstable dynamically, unless the source was significantly in front or behind RW Aur. We additionally note that there are other similarly sized clumps of CO emission seen in the moment maps. Therefore, if the third CO peak is tracing a third stellar or substellar companion in the RW Aur system, it would have an unusually low amount of millimeter-sized dust and would be on an unstable orbit.  Another possible explanation would be that the third CO peak is connected to the $\gamma$ stream by a fainter unseen structure. Complementary (u,v) coverage at intermediate baselines would be necessary to confirm or exclude this possibility.



\begin{figure*}[!ht]
\centering\includegraphics[width=0.9\linewidth, trim = 0 3.2in 0 0]{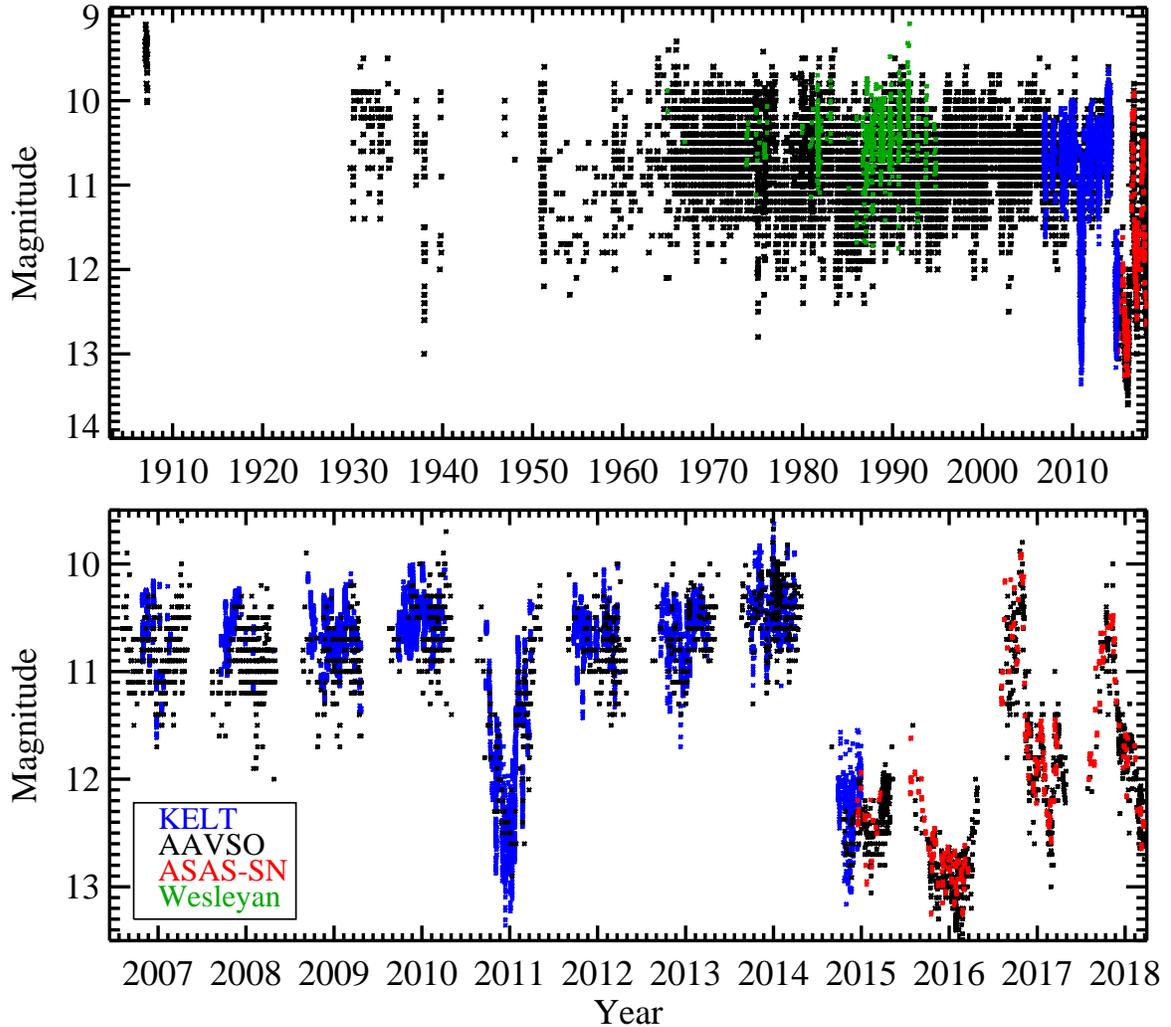}
\caption{Recreation of Figure 1 from \citet{Rodriguez:2013, Rodriguez:2016A} (Top) showing the full $\sim$110 year long lightcurve from AAVSO (Black), Wesleyan (Green), KELT (Blue), and ASAS-SN (Red) of RW Aur and (Bottom) a zoom in showing the photometric variability of the system over the last decade.}
\label{figure:LC}
\end{figure*}

\begin{figure}[!ht]
\centering\includegraphics[width=1.0\linewidth, trim = 0cm 0cm 0cm 0cm]{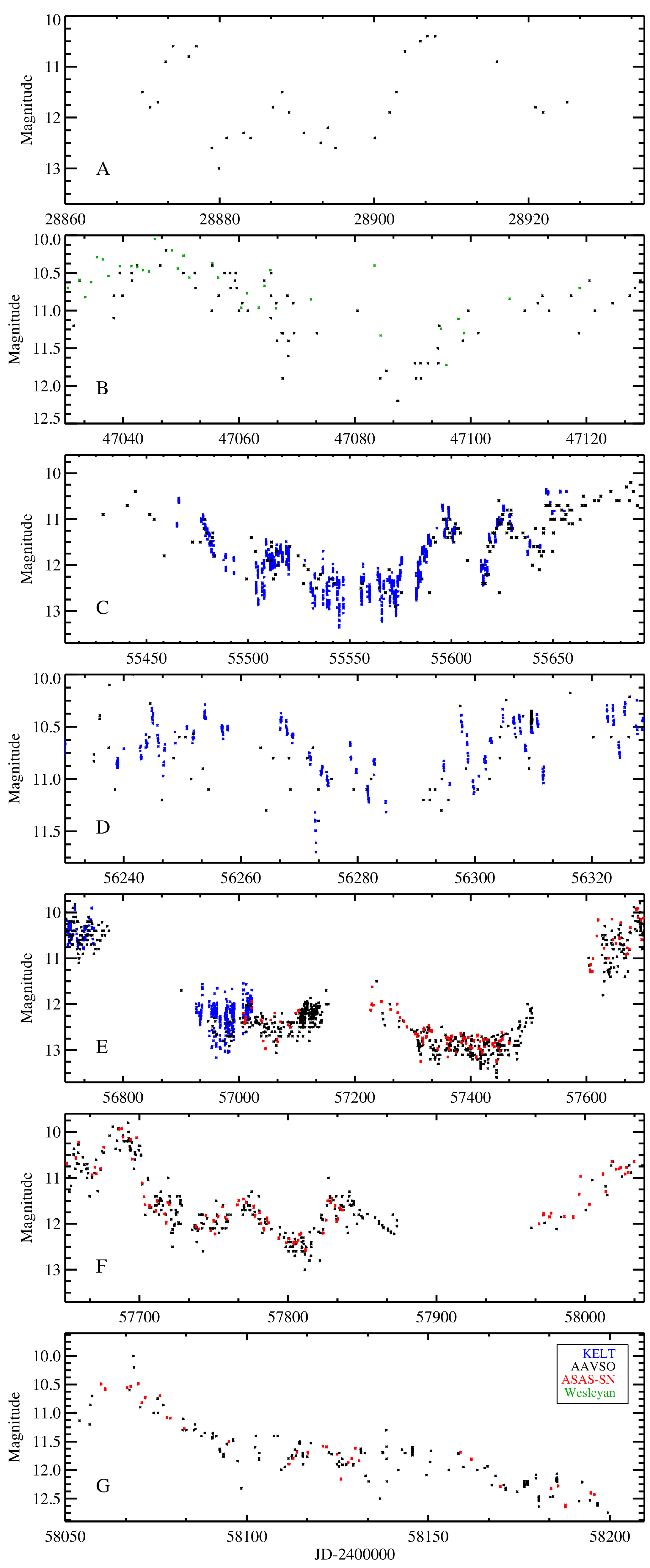}
\caption{All seven dimmings of RW Aurigae: (A) Archival observations showing the 1937-1938 dimming, (B) The 1987-1988 dimming first identified by \citet{Berdnikov:2017}, (C) Recreation of Figure 3 from \citet{Rodriguez:2013} showing the 2010-2011 event, (D) Recreation of Figure 3 from \citet{Rodriguez:2016A} showing the 2012-2013 event, (E) The full 2014-2016 dimming. (F) The Full 2016-2017 dimming. (G) The dimming that began in late 2017.}
\label{figure:Eclipses}
\end{figure}

\section{Photometric Results}
\label{Phot}
Over the past decade the RW Aurigae system has undergone four separate dimming events that have varied in duration and depth \citep{Rodriguez:2013, Petrov:2015, Antipin:2015, Rodriguez:2016A, Bozhinova:2016, Berdnikov:2017}. We report here a new dimming event that is currently on-going, as well as analyzing an event dating back to 1937 that was first identified by \citet{Berdnikov:2017} (see Figure \ref{figure:LC} \& \ref{figure:Eclipses}). The cause of these dimmings is debated but may be due to a dusty disk wind very close to the star \citep{Petrov:2015, Shenavrin:2015}, disrupted circumstellar material from the eccentric fly-by farther out \citep{Rodriguez:2013, Rodriguez:2016A}, variation in the geometry of the inner disk \citep{Schneider:2015, Facchini:2016} similar to what has been proposed to explain the large dimmings of AA Tau \citep{Bouvier:2013, Loomis:2017}, or tidal perturbations of the RW Aur A disk propagating inward since the flyby and now reaching the inner 1 AU \citep{Berdnikov:2017}. 

In this section, we summarize the eclipse parameters for each of the seven dimmings, including one that occurred in the late 1930s. We model each event separately as an occultation of RW Aur A by large opaque screen with a leading edge perpendicular to its direction of motion (following \S5.1 in \citealp{Rodriguez:2013}). From this analysis, we estimate the transverse velocity and width of the occulting body required to explain each dimming event separately and summarize the results in Table \ref{tbl:phot}. The analysis in this section provides an estimated range of the size and velocity of the occulting dust screens to explain each event. In each dimming, we see in-eclipse photometric structure which we attribute to changes in the opacity and/or vertical size of the occulting bodies. Therefore, our estimated velocities are likely a lower limit since we are assuming a sharp leading edge of an opaque occulting screen when the observed structure during the eclipse would suggest otherwise. The lower limit on the velocity results in an upper limit on the estimated semi-major axis. The quoted capitalized letters in the following sub-section headers correspond to the same letter label shown in the bottom left of each panel in Figure \ref{figure:Eclipses}. In \S\ref{photsummary}, we summarize our results and discuss them in context with our analysis of the ALMA observations.

\subsection{Historical Dimming: 1937-1938 ``A'', 1987 ``B''}
Prior to the 2010-2011 dimming, there had been no similar event observed in the previous $\sim$60 years. Although the baseline brightness of RW Aur has varied over the past 110 years \citep{Berdnikov:2017}, there is no gap in the AAVSO observation large enough that a 180 day event of similar depth would have been missed since 1960 (see Figure \ref{figure:LC}, \citealp{Rodriguez:2013}) and no similar (depth and duration) event observed over the last $\sim$110 years \citep[e.g.,][and references therein]{Berdnikov:2017}. However, two shallower and shorter episodes of dimmings were observed in 1937 and 1987. In early December of 1937, RW Aur dimmed by $\sim$2 mag for $\sim$30 days, ending in early January of 1938. During this dimming, RW Aur appears to brighten by 1 magnitude for $\sim$10 days during the middle of the event (see Figure \ref{figure:Eclipses}). The initial dimming of this event was not observed in our $V$ band observations but RW Aur takes $\sim$6 days to recover to its normal brightness. However, the initial dimming was observed in the $B$ band showing it to be similar in duration. Prior to the 1937-1938 event, a very short but similar depth dimming event was seen in the $B$ band observation \citep{Berdnikov:2017}. Additionally, there is also short duration photometric variability seen after this event, consistent with trailing circumstellar material to the main occultating screen. This dimming occurred during a known 10-15 year long change in the nominal brightness of the RW Aur system \citep{Berdnikov:2017} but we note that the depth, shape, and short duration of the event argues that this is separate to the known baseline brightness change. Additionally, \citet{Berdnikov:2017} identified a dimming event that began in late UT 1987 September and ended a few months later in UT 1987 November. From reanalyzing observations from the AAVSO and Weleyan archive, we determined that this event had an ingress timescale of $\sim$30 days, very similar to what was seen for the 2010-2011 event. The dimming lasted about 50 days and reached a maximum depth of $\sim$1.2 mag. Following the same model for these two dimmings as done for the 2010-2011 and 2012-2013 events and adopting a stellar radius of 1.6$\pm$0.32 \rsun \citep{Rodriguez:2013}, we estimate the transverse velocity of the occulting bodies to be $\sim$4.6 (1937-1938) and 0.9 \kms (1987), with both events yielding a similar occulting body width of $\sim$0.08 and 0.03 AU.



\subsection{2010-2011, 2012-2013, 2014-2016 Dimmings ``C, D, \& E''}
From analyzing photometric observations from the KELT survey and the AAVSO, the eclipse in 2010-2011 had a duration of $\sim$180 days and a depth of $\sim$2 mag in the optical. A few years later in 2012-2013, RW Aur experienced a 0.7 mag dimming event for $\sim$40 days. Using kinematic and geometric arguments, it was determined that the occulting bodies would be moving at 0.9--2.7 \kms, corresponding to a $\sim$0.3 and $\sim$0.06 AU in width for the 2010-2011 and 2012-2013 events, respectively \citep{Rodriguez:2013, Rodriguez:2016A}. Unfortunately, these dimmings were not discovered until RW Aur had recovered back to its nominal brightness, hindering our ability to understand these phenomena through targeted multi-band photometric and spectroscopic observations.

From the continued monitoring effort of RW Aur by the AAVSO, the RW Aur system appeared significantly dimmer after the mid-2014 observing gap \citep{Petrov:2015}. Combining these observations with additional data from KELT and the Kutztown University Observatory (KUO), it was determined that the RW Aur system had once again dimmed by $\sim$2 mag and remained in a dim state through the 2014-2015 observing season \citep{Rodriguez:2016A}. Spectroscopic observations prior to and during this dimming showed no changes in the emission of H$\alpha$ or He I line, suggesting that the high accretion rate for RW Aur A remained constant. Resolved photometric observations of RW Aur showed that RW Aur A was $\sim$3 magnitudes fainter than the RW Aur B in all optical filters \citep{Lamzin:2017}. Additionally, infrared photometric monitoring showed a clear decrease in the $JHK$ brightness of the system but an apparent increase in the $M$ and $L$ filters \citep{Shenavrin:2015}. This IR excess was explained by $\sim$1000 K dust only 0.1--0.2 AU from RW Aur A and is likely associated with this dimming event. Due to the seasonal observing gap, we are unable to determine precisely when this event began. In late March of 2016, RW Aurigae began to brighten prior to the 2016 observing gap. After the 2016 observing gap, RW Aur was at a brightness of V$\sim$10.8. Unfortunately, both ingress and egress for this event occurred during seasonal observing gaps, limiting our ability to estimate the transverse velocity of the the occulting body. We estimate a total duration of the 2014-2016 dimming to be $<$830 days, assuming the first point after the 2016 observing gap as the end of the eclipse.

\subsection{Recent Events: 2016-2017 ``F'', 2017-2018 ``G''}
In late October of 2016, after about $\sim$84 days from the end of the 2014-2016 eclipse, RW Aurigae dimmed by $\sim$2 mag for the fourth time in less than a decade. This dimming was first identified by \citet{Berdnikov:2017}. We estimated the ingress to be $\sim$35 days and the egress to be $\sim$65 days. Using the same model, this would correspond to a transverse velocity range of 0.4--0.8 \kms. The event lasts until early October of 2017, with a duration of $\sim$335 days. Similar to what was seen in the 2010-2011 and 2012-2013 dimmings, the 2016-2017 dimming displays a sinusoidal variation on a similar timescale as the ingress/egress. We estimate the width of the occulting body to be $\sim$0.08--0.15 AU. The longer estimated egress timescale may indicate that the occulting body has a sharp leading edge but a diffuse, more extended trailing edge. 

After RW Aurigae recovered from the 2016-2017 dimming, it only stayed at its nominal brightness for $\sim$82 days before beginning a fifth dimming event in less than a decade. Since the event is ongoing, we only have a small amount of photometric coverage.The dimming event has been going in for $>$120 days. Although it appears that RW Aur is still dimming, it is possible that the ingress is $\sim$25 days beginning at 2458067 and ending on 2458093. This would suggest a similar ingress timescale as the 2013 dimming event. To date, this dimming has reached a depth of $\sim$2 mag but the system may continue to fade.


\begin{table*}
 \centering
 \caption{}
\scriptsize
\label{tbl:phot}
 \begin{tabular}{lccccccc}
    \hline
    \hline
ID&    Dimming Event & Ingress/Egress & Transverse Velocity & Duration & Occulter Width & Maximum Depth& Discovery Reference \\
    \hline
&& (days)& \kms & (days) &(AU) & (mag)\\
    \hline
A&1937--1938 & $\sim$6 & 4.6 & $\sim$30 & 0.08 & $\sim$2& \citet{Berdnikov:2017} \\ 
B&1987 & $\sim$30 &0.9 & $\sim$50 & 0.03 &$\sim$1.2& \citet{Berdnikov:2017} \\    
C&2010--2011 & 10-30 & 0.9--2.7 &$\sim$180 & 0.1--0.3 & $\sim$2& \citet{Rodriguez:2013}  \\
D&2012--2013 & $\sim$10 & 2.7 & $\sim$40 & 0.06 & $\sim$0.7 & \citet{Rodriguez:2016A} \\
E&2014--2016 & --- & --- & $<$830 & --- & $\sim$2.5& \citet{Petrov:2015, Rodriguez:2016A}\\
F&2016--2017 & 35--65 & 0.4--0.8 & $\sim$335 & 0.08--0.15 & $\sim$2& \citet{Berdnikov:2017} \\
G&2017--?    &  $\ge$25 & ? & $>$120 & ? & $\ge$2& this work \\
\hline
 \end{tabular}
\begin{flushleft} 
 \footnotesize{ \textbf{\textsc{NOTES: The methodology in estimating the kinematics of the occulting body are described in \S\ref{Phot}.}
}}
\end{flushleft}
\end{table*} 

\section{Discussion}
\label{discussion}

\subsection{Summary from over a Century of Photometric Observations}
\label{photsummary}
From analyzing over 100 years of photometric observations, there have been 7 separate dimming events identified. It is likely that the eccentric fly-by of RW Aur B, that disrupted the disk around RW Aur A, has significantly influenced the circumstellar environment contributing to the observed dimming events. From applying a simple model and analyzing each event independently, we estimate physical parameters of the occulting screens which are shown in Table \ref{tbl:phot}. For each event, we derive a transverse velocity in the range of 0.4 to 4.6 \kms.
Using the 1937-1938 event (which has the fastest estimated velocity) and a 1.4\msun for RW Aur A, this would correspond to a semi-major axis of $\sim$55 AU, assuming Keplerian motion. This would place the occulting body outside the radius of the dust ($\sim$21 AU, see \S\ref{continuum}) and gas ($\sim$58 AU, see \S\ref{SpectralLine}) disks, where Keplerian rotation speeds would be respectively 8 \kms\ and 4.7 \kms\ (around a 1.4 \msun\ star). This may suggest that the occulting material causing this event may not be at the same orbital distance as the material that have caused the more recent events.

One change to our analysis that could allow the occulting body to be in orbit within the short truncated disk around RW Aur A would be to have an inclined leading edge relative to the direction of motion (see Figure 9 from \citealp{Rodriguez:2015}). This model allows one to increase the occulting body's velocity, placing it closer to the host star, assuming Keplerian motion. Also, this inclined edge geometry is what one would expect for the shape of a warp within the disk.  Using the 1937-1938 event, which had the shortest ingress timescale (and thus the largest estimated velocity), it would take a leading edge angle of 60.5$^{\circ}$ to place the occulting body at the edge of the dust disk. For the 2016--2017 event, which had the longest ingress/egress timescale, it would require a leading edge angle of 4.5$^{\circ}$. 

We also note that the disk radius in the mm range is dominated by large grains; hence, smaller dust grains could still be present up to the edge of the gas disk (at 58 AU), and further out in stripped material off the disk plane. We note that the estimated velocity derived from each dimming event is consistent with the relative velocities of the observed disrupted gas (see Figure \ref{figure:Band6CO}). Given the evidence of a star--disk interaction, that has significantly disrupted the circumstellar environment around RW Aurigae, we are not able to place any additional constraints on the location of the occulting features. 

The large range in the duration of the seven dimming events, resulting in a wide range of occulting body sizes, could be explained by multiple dust clumps orbiting RW Aur A in Keplerian motion \citep{Rodriguez:2013}. However, the large variations in the occulting body parameters could also easily be explained if the dust screen is caused by disk winds \citep{Petrov:2015} or inner disk perturbation \citep{Facchini:2016}. Since the estimated kinematics for the more recent events are similar (a few \kms, we note that this is not clear for the 2014-2016 event), it is very likely that they are all caused by the same mechanism. Unfortunately, we are not able to distinguish between the plausible interpretations presented in the beginning of \S\ref{Phot}. 

There is a $\sim$73 year baseline between the start of the 1937 and the start of the 2010 dimming events. Interestingly, Photometric dimmings were observed shortly before and after the 1937 event \citep{Berdnikov:2017}. Therefore, it is possible that these dimmings occur periodically, but in clusters of dimming events. However, we see only the one dimming event in the late 1980s where our photometric coverage is complete enough to not have missed additional dimmings. Taking the full baseline between the first and last dimming event ($\sim$73 years) and assuming Keplerian motion around a 1.4\msun star (RW Aur A), this would correspond to a semi-major axis of $\sim$20 AU. Using the full 110 year baseline for our photometry (see Figure \ref{figure:LC}), we are able to search for periodic phenomena that correspond to a spatial scale of $\sim$26 AU. This shows that the long baseline of photometric observations from networks like the AAVSO are beginning to probe the same spatial scales as our ALMA observations.

\subsection{Nature and Morphology of the Tidal Streams}
The presence of additional tidal streams would not be explained with a single star--disk encounter  \citep{Clarke:1993, Munoz:2015}. One explanation is that RW Aur B had an initial circumstellar disk prior to the eccentric fly-by, and we are seeing the aftermath of a disk-disk collision. Even if RW Aur B had an initial disk, it is unlikely we are seeing the aftermath of one random encounter. The probability for an unbound star--disk fly by would $\sim$10$^{-4}$ per star per Myr (see Table 1 of \citealp{Clarke:1991}), making it very unlikely that this is a random encounter. Additionally, the probability is even less since both RW Aur A and B are known to be pre-main sequence stars \citep{Duchene:1999}, show strong Li I 6707 absorption lines \citep{StoutBatalha:2000, Takami:2016}, and are located near the outer edge of the Taurus-Auriga star formation region. Therefore, we propose that RW Aur A has undergone multiple eccentric fly-bys ($\ge$2) from RW Aur B, significantly disrupting the circumstellar material in the system. 

\begin{figure}[!ht]
\vspace{0.3in}
\centering\includegraphics[width=0.95\linewidth, trim = 1in 0.0in 1in 1in, angle = 0]{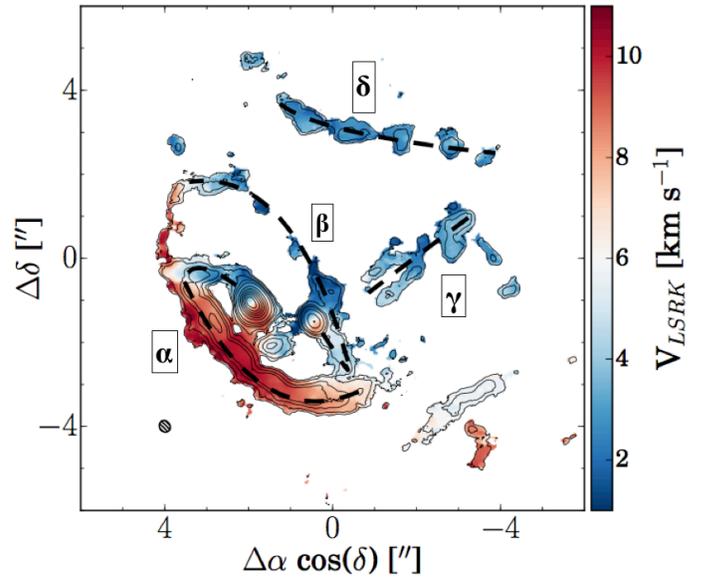}
\caption{Moment-1 map as shown in Figure \ref{figure:Band6CO} overlaid with moment-0 contours. The original tidal arm discovered by \citet{Cabrit:2006} is labeled ``$\alpha$", the second counter spiral arm  is labeled ``$\beta$''and the two new tidal streams identified are labeled ``$\delta$" and ``$\gamma$". The dashed black line outlines the two spiral arms that form the apparent ring and the additional tidal streams to the NW. }
\label{fig:mom1_ann}
\end{figure}

From the new high-resolution ALMA observations, we now know that the original tidal arm discovered by \citet{Cabrit:2006} (``$\alpha$") appears to be part of a projected circumbinary ring of gas around both RW Aur A and B. The observed blue-shifted material seen connected to the NW of RW Aur B's disk (``$\beta$") is consistent with a second counter-spiral arm to the original tidal arm. Figure \ref{fig:mom1_ann} outlines the two spirals that would form the projected ring. Therefore, we may be observing the second tidal arm of gas that is connected to the NW extent of the disk around RW Aur B, as suggested from the SPH simulations (see Figure 11 in \citealp{Dai:2015}). 

To distinguish between the dual spiral arms and circumbinary ring scenarios, we calculate the escape velocity for the arm and compare this to the observed relative velocity in the $^{12}$CO(2-1) moment 1 map (see Figure \ref{figure:Band6CO}). We estimate that the arm ranges from 0.95 -- 3$\arcsec$ (130 -- 420 AU) from RW Aur A and 1.95 -- 3$\arcsec$ (273 -- 420 AU) from B. Using a mass of 1.4\msun and 0.9\msun for RW Aur A and B \citep{Ghez:1997}, this corresponds to an escape velocity of 2.4 -- 4.4 \kms and 1.9 -- 2.4 \kms for RW Aur A and B, respectively. From the moment 1 map, the arm (``$\alpha$'') is red-shifted relative to RW Aur A by up to 6\kms, with the largest redshift component located in the center of the arm, as seen by \citet{Cabrit:2006}. This suggests that the arm is unbound at the center and possibly partially bound at the NE and SW extensions, where the arm is closest to either RW Aur A or B. This provides evidence that the apparent ring is likely formed from the original tidal arm connected to the RW Aur A disk and a second counter-spiral arm connected to the RW Aur B disk, as suggested from the SPH simulations \citep{Dai:2015}. We note that the measured redshift and relative distances from RW Aur A and B are lower limits, as we have not corrected for line of sight projection effects. A comparison between the calculated escape velocity and observed relative velocity suggests that two spirals are partially unbound and likely expanding outward, as suggested by \citet{Cabrit:2006}.

In S\ref{tidalstreams}, we identified two additional tidal streams ``$\gamma$" and ``$\delta$", outlined in Figure \ref{fig:mom1_ann}, to the first discovered tidal arm and its counter spiral (``$\alpha$" and ``$\beta$").  The ``$\gamma$" stream extends out $\sim$4$\arcsec$ in the NW direction from RW Aur B. The escape velocity at this separation from RW Aur A ($\sim$5.6$\arcsec$) and B ($\sim$4.4$\arcsec$) is 1.8 and 2 \kms, respectively. The stream is blue shifted by 1-3 \kms relative to the motion of the entire system suggesting that some of the tidal stream in unbound. The fourth tidal stream ``$\delta$" in the far NW 
(see Figure \ref{fig:mom1_ann}) is separated from RW Aur A by 4.3--6.3$\arcsec$ (602--882 AU) and is blueshifted by 2-3 \kms. The estimated escape velocity is 1.7 \kms. Therefore, this fourth tidal stream is likely completely unbound to the system and expanding away.  It is not clear at this stage how to reproduce the morphology of the tidal streams ``$\delta$" and ``$\gamma$" even with multiple encounters. Further modelling, taking into account an improved orbital solution and a possible secondary disk around RW Aur B prior to the first fly by, are required.

\subsection{Gas Properties in the Tidal Arm and Disks}
\label{sec:LVG}
\begin{figure}[!ht]
\vspace{0.3in}
\centering\includegraphics[width=0.99\linewidth]{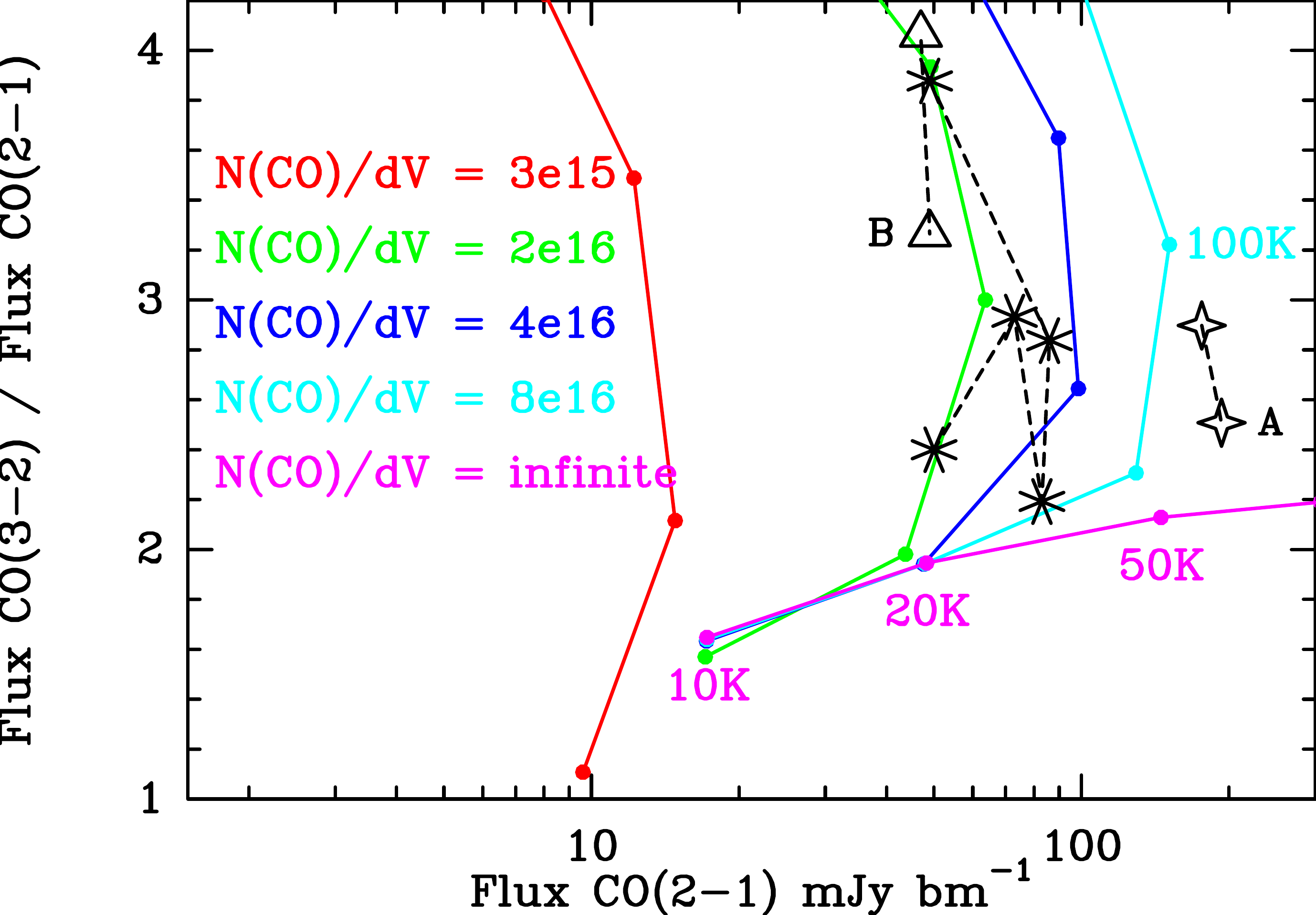}
\caption{The $^{12}$CO(3-2)/$^{12}$CO(2-1) flux density ratio as a function of the peak line flux density in $^{12}$CO(2-1) measured towards the blue and red sides of the RW Aur A disk (connected stars), the blue and red sides of the RW Aur B disk (connected triangles), and five positions along the main tidal arm (connected asterisks, see text). Coloured curves show LVG predictions in LTE for an isothermal region with various values of opacity parameter} $N$(CO)/$dV$ (as labelled, in cm$^{-2}$ per \kms) and gas temperatures of 10, 20, 50 and 100~K (dots along each curve) increasing from lower left.
\label{fig:co32co21}
\end{figure}

To estimate the range of gas temperatures $T_k$ and column densities in the CO-emitting region of the tidal arm, we plot the observed $^{12}$CO (3-2)/(2-1) ratio as a function of the surface brightness at the line peak in $^{12}$CO (2-1) at various positions (see Figure \ref{fig:co32co21}) and compare to simple predictions for an isothermal slab in the Large Velocity Gradient (LVG) approximation, calculated with the on-line RADEX modeling tool \citep{radex}. We assume densities high enough to be in LTE, so as to have only two free parameters left: the gas kinetic temperature $T_k$ and the column density per unit velocity, $N$(CO)/$dV$. We show in Figure \ref{fig:co32co21} five positions along the arm, including the portion of the arm that is connected to the NE extension of the RW Aur A disk (which we refer to as the ``crook"), the 3 emission peaks along the brightest portion of the arm (near ``$\gamma$''), and the ``end" of the arm at (-1.95\arcsec,-2.45\arcsec). For comparison, we also plot the same parameters observed at the line peaks of the A and B disks (both on their blue- and red-shifted sides). 

We first note that we can readily obtain a lower limit to $T_k$ by comparing the $^{12}$CO (2-1) surface brightness to the predicted values in the optically thick limit (magenta curve). We obtain $T_k \ge 20-30$ K in the arm, slightly larger than the 10 K estimated by \citet{Cabrit:2006} from PdBI maps at lower angular resolution, suggesting some degree of clumpiness. In the B disk, we find $T_k \geq 20$~K. We note that any beam-dilution correction would shift the observed points to the right in Figure \ref{fig:co32co21} (the line ratio would be unaffected). Hence these are solid lower limits to the true $T_k$. In the A disk, which is likely optically thick in CO and appears well resolved, we find $T_k \simeq 70$~K, in excellent agreement with earlier estimates of 60-110~K obtained by \citet{Cabrit:2006} from simplified CO line profile modeling. 

We also note that the ratio of $^{12}$CO(3-2)/(2-1) is always slightly above the predicted value $\simeq 2$ for optically thick isothermal emission, by a factor 1.5 to 2. The curves in Figure \ref{fig:co32co21} show that in the isothermal assumption, this would require marginally thin emission at higher temperature than in the optically thick limit: assuming negligible beam dilution, $T_k$ in the arm would be 100 K at the crook and 25-50 K elsewhere. An alternative explanation for the ratios $> 2$ would be that the emission is optically thick and non isothermal, with the $\tau = 1$ surface of CO(3-2) probing 1.5-2 times hotter gas than the $\tau =1$ surface of CO(2-1). This is the most likely explanation for the ratios $\simeq 2.5-4$ observed in the A and B disks (Fig \ref{fig:co32co21}), since CO emission remains optically thick out to radii of 100-200 AU in typical T Tauri disks, with CO(3-2) being thicker than CO(2-1) and probing higher and warmer atmospheric layers. The same situation may apply to the tidal arm if it has retained some of the disk's internal temperature gradient after tidal stripping, or if it is externally heated. An upper limit on the CO optical depth in the arm can be derived from the 3$\sigma$ upper limit in $^{13}$CO(2-1) at the peak of the arm (5.5 mJy/beam in a beam of 0.27" $\times$ 0.23"); we find a flux density ratio of  $^{12}$CO(2-1)/$^{13}$CO(2-1)$> 13$ (after correction for the difference in beam sizes). Assuming similar excitation in the two isotopologues and a solar isotopic abundance ratio of 89, we infer a $^{12}$CO(2-1) optical depth at the peak $< 7$. Hence optically thick and non-isothermal CO in the arm at 20-40 K is not ruled out.





The isothermal, optically thin case still allows us to set a lower limit to the CO column densities, which in the arm would range over $N$(CO)/$dV$ $\simeq 2-8\times 10^{16}$ cm$^{-2}$/(\kms). Using the determined typical line width in the arm of 2 \kms, we obtain a minimum CO column density of $N_{\rm CO} \geq 4-16 \times 10^{16}$ cm$^{-2}$. Since some CO may still be locked in ice mantles, we expect a gas-phase abundance [CO]/[H] $\le 10^{-4}$ (the typical ISM value), and infer a lower limit to the hydrogen column density of $N_{\rm H} \ge 0.4-1.6 \times 10^{21}$ cm$^{-2}$. From this, it appears that a portion of tidal arm passing on the line of sight would contain enough dust to dim the star. We also infer that the arm is a very dense structure. As mentioned in Section 3.2.2, the arm has a typical width of $\simeq 85$ AU. Hence we obtain a lower limit to its (mean) volume density, $n_H \ge 3-12 \times 10^5$ cm$^{-3}$. This confirms \textit{a posteriori} that LTE should be valid (the critical densities of the CO(2-1) and (3-2) lines at $\simeq 20$K are $2\times 10^3$ and $\simeq 7 \times 10^4$ cm$^{-3}$, respectively).

However, we stress that these are only strict lower limits to $N_H$ and $n_H$, as CO freeze-out at high densities could make a large fraction of the gas invisible. If we were to neglect CO freeze-out and adopt a standard CO gas-phase abundance of $10^{-4}$ in the arm, the CO(2-1) flux integrated over the arm of 3.7 Jy \kms\ would translate into a total gas mass of $5 \times 10^{-6} M_\odot$, where the conversion is for optically thin CO emission in LTE with $T_k$ in the range 10--50~K \citep[see Section 3.2.3 in][]{Cabrit:2006}. Correcting for a maximum optical depth of $\tau_{12} \le 7$ would increase the mass up to $3.5 \times 10^{-5} M_\odot$, which is still a very low value. In their analysis of the IM Lup disk, \citet{Pinte2018} found that the gas-phase CO abundance drops by a factor $10^{-4}$ below 21 K, so that none of the cool disk midplane is detectable. Applying the simple mass formula above to the CO(2-1) flux of IM Lup (24.7 Jy \kms), we would infer a disk mass of $2.5 \times 10^{-4} M_\odot$, whereas the true disk mass in the illustrative model of \citet[][Appendix B]{Pinte2018} is 0.6$M_\odot$. Because of this, it is not possible to estimate accurately the total gas disk mass from CO lines alone \citep{Long2017}. The fact that CO excitation temperatures in the tidal arm range down to $\simeq 20$ K in Figure \ref{fig:co32co21}, which is close to the freeze-out temperature of 21 K determined by \citet{Pinte2018}, strongly suggests that CO emission in the arm may also be "freeze-out limited" and suffer from the same problem as in disks. 

Keeping these caveats in mind, we note that under the ideal case where (1) CO ice mantles on dust grains have not been sputtered back into the gas phase during tidal stripping (not unreasonable given the low expected shock speeds $<$ 5 \kms), and (2) the structure in density and temperature is not too different between CO emitting regions in the tidal arm and in the outer disk of A (again not impossible since the CO surface brightness and line ratios are similar in Fig. \ref{fig:co32co21}), then the ratio of CO line fluxes between arm and disk would directly yield the mass ratio of gas that was stripped off in the arm compared to that left in the A disk after truncation, without requiring us to know the exact amount of CO depletion (it would cancel out in the ratio). With integrated flux densities of 3.4 Jy \kms in the A disk and 3.7 Jy \kms over the arm, this ratio would turn out to be close to 1, and the mass of the ejected material would then be $\sim 50\%$ of the initial mass of the disk. The tidal encounter would thus have been able to remove a large amount of the disk mass, and likely most of the angular momentum. Interestingly, \citet{Munoz:2015} have shown that when the pericentre of a star-disk encounter is comparable to the disk radius, the gravitational interaction can extract a significant amount of the angular momentum of the disk thus favouring stellar capture, provided that the disk is massive. If the two stars are bound, the undergoing tidal encounter would then have likely removed orbital energy from the system, making the system more bound, and reducing the orbital period. However, we stress that the ejected mass ratio of 50\% derived above is obtained under a very restrictive set of assumptions, and is very uncertain. Numerical simulations of the encounter with improved orbital parameters would be necessary to clarify the past and future orbital evolution of the system.

\subsection{Tidal Truncation of the Disks}
The apparently complete encirclement of RW Aur by ejected material suggests that the system has been subject to more than one encounter - i.e it is a bona fide binary, rather than a one off encounter. To estimate the tidal truncation of the primary by the secondary (and vice versa) in this case we follow \cite{2005MNRAS.359..521P}, \cite{Harris2012}, where the truncation radius of the primary is 
\begin{equation}
	R_t \approx 0.337 \left[ \frac{\left(1-e\right)^{1.2} \psi^{2/3}  \mu^{0.07}}{0.6\psi^{2/3} + \textrm{ln}\left(1+\psi^{1/3}\right)} \right] \mathcal{F} a_p
    \label{equn:truncationRadius}
\end{equation}
and $e$, $\psi$, $\mu$ and $a_p$ are
the eccentricity, mass ratio of the primary star to companion ($\psi=M_p/M_s = 1/q$), the mass fraction of the stellar pair ($\mu = q/\left[1 + q\right]$) and the projected separation respectively. $\mathcal{F}$ is the ratio of semimajor axis to projected separation, which \cite{Torres:1999} show is
\begin{equation}
	\mathcal{F} = \frac{a}{a_p} = \frac{1}{1-e\cos(E)} \frac{1}{\sqrt{1-\sin^2\left(\omega+\nu\right)\sin^2 i}}
\end{equation}
where $a$, $\omega$, $\nu$, $E$ and $i$ are the actual separation, longitude of periastron, true anomaly, eccentric anomaly and inclination respectively. Inverting the stellar mass ratio gives the truncation of the secondary disk.

Given the unknowns in the above equations, we follow \cite{Torres:1999}, \cite{Harris2012} and use a Monte Carlo approach to construct a probability distribution for $R_t$. That is we set up 1000 bins in radius, spanning 0 to 400AU. We then perform 10 million random samplings of the unknown orbital parameters and place the resulting $R_t$ in the appropriate bin for each random realization. Normalizing the resulting binned distribution gives the probability $P(R_t)$. 

The randomly sampled parameters are as follows. The longitude of periastron is sampled uniformly from 0 to $2\pi$ and eccentricity from 0 to some upper limit (we explore both 1 and 0.7). The inclination has a sinusoidal dependence so is sampled evenly in $\textrm{asin}(r)$ for random variable $r$ in the range 0:1. Lastly, we require random eccentric and true anomalies. The mean anomaly $M$ is constant with orbital phase and so is randomly sampled with uniform probability from 0 to $2\pi$. This is related to the eccentric anomaly by
\begin{equation}
	M = E - e\sin(E)
\end{equation}
from which we solve for $E$ by bisection from E=0 to 2$\pi$. The true anomaly is then
\begin{equation}
	\nu = 2\, \textrm{atan}\left[\sqrt{\frac{1+e}{1-e}}\tan\left(\frac{E}{2}\right)\right].
\end{equation}


\begin{figure}
    \centering
    \includegraphics[width=9cm]{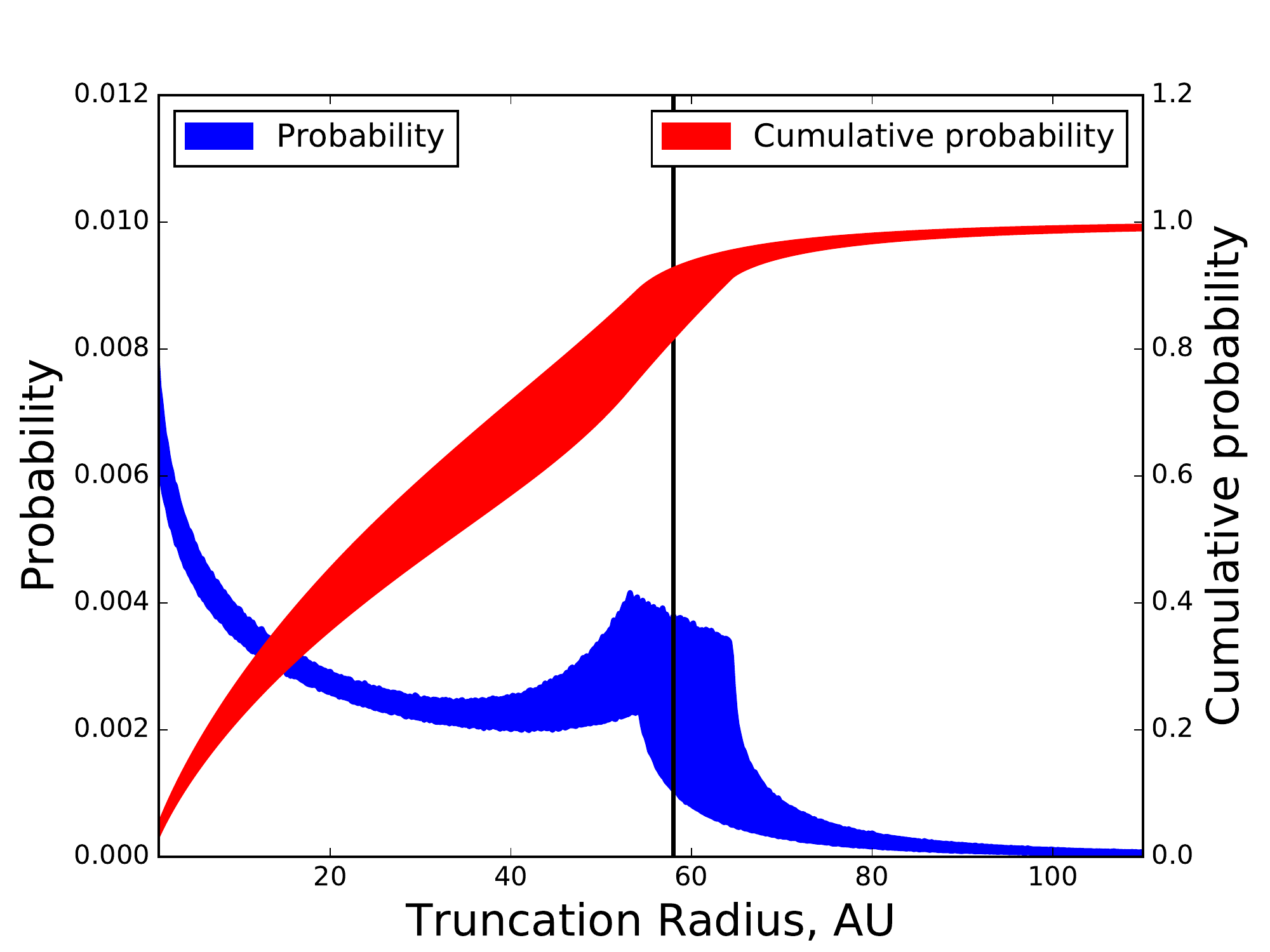}
    \includegraphics[width=9cm]{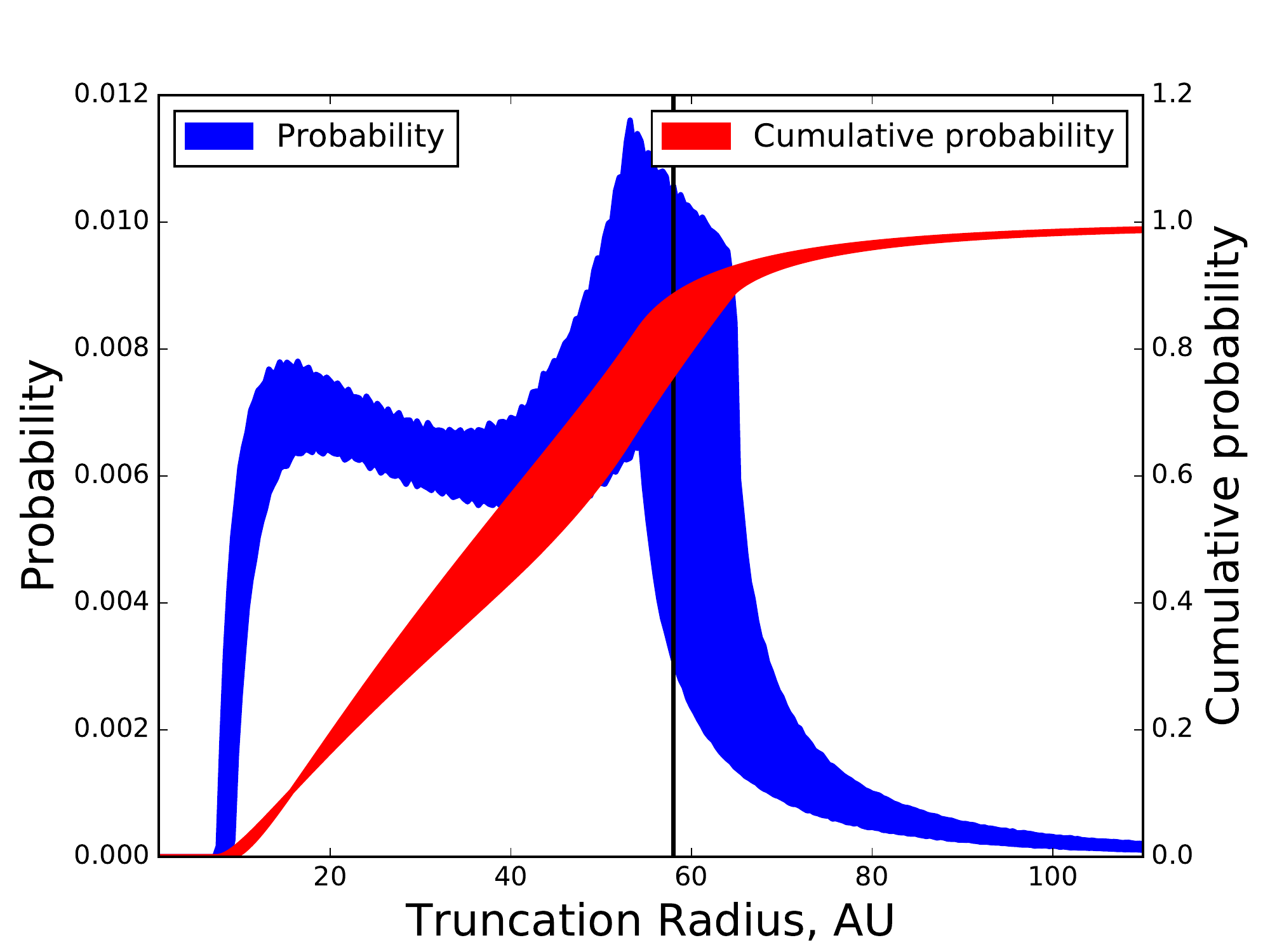}
    \includegraphics[width=9cm]{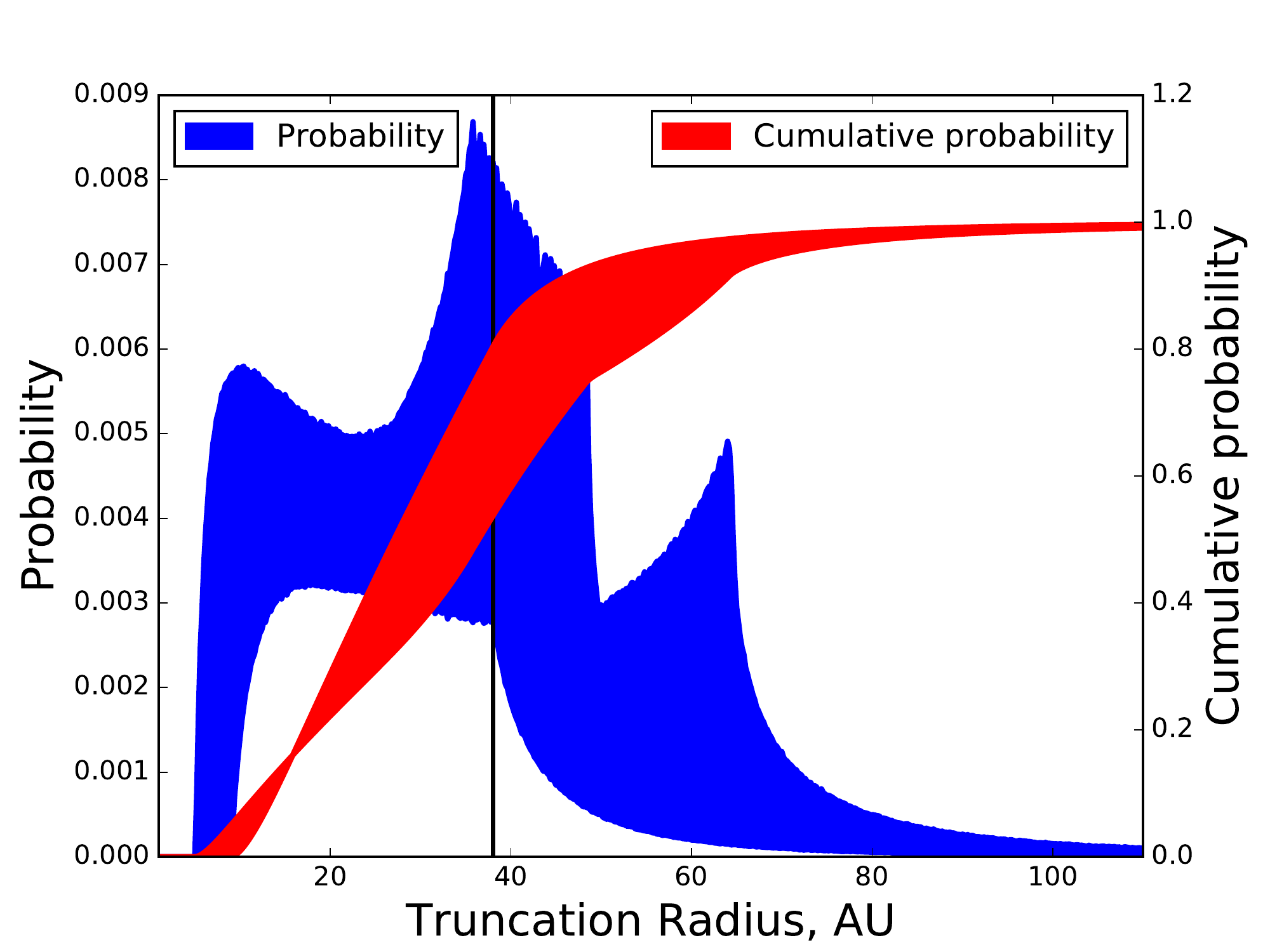} 
    \caption{The (cumulative) probability distribution of truncation radius of the RW Aur disks. The multi-valued nature of the distributions at any given truncation radius is due to the uncertainty in the stellar masses. The upper and middle panels are the distributions for the primary disk, with a maximum eccetricity of 1 and 0.7 respectively. The lower panel is the probability distribution for the secondary disk for a maximum eccentricity of 0.7. The black vertical line in each case is the current observed extent of the disk. }
    \label{fig:Truncation}
\end{figure}

For the current observed separation of $1.468\pm0.056\arcsec$, distance of 140\,pc, and possible stellar masses of 1.3-1.4 and 0.3-1.0\,M$_\odot$ for the primary and secondary respectively, the truncation radius probability distribution is shown in Figure \ref{fig:Truncation}. The upper panel is for the primary disk with a maximum eccentricity of 1 and the middle panel for a maximum eccentricity of 0.7. The lower panel is for the secondary disk with a maximum eccentricity of 0.7. The multi-valued nature of the distribution comes from the spread in possible stellar mass ratios.

Both the 58 and 38 AU observed gas radii of the primary and secondary disks (Table \ref{tbl:Results}) are consistent with the most probable tidal truncation radii in Figure \ref{fig:Truncation} for an eccentricity of 0.7, (though note that we have not constrained the eccentricity through this approach). Although evidence of interaction is obvious in RW Aur in the form of ejected material, this provides further evidence that the disk extents are being set by the gravitational interaction of the binary pair. Furthermore, since an eccentricity $<1$ gives a peak in the distribution consistent with the observed disk radius, this is further evidence for the system being bound and having undergone a number of interactions.

\section{Conclusions}
\label{conclusion}
The main results from our analysis of the new ALMA observations of RW Aur combined with the $\sim$100 years of photometric observations are the following: 

\begin{itemize}
    \item Both RW Aur A and B show clear evidence of Keplerian motion in the disks around each star. RW Aur A is $\sim$8 times brighter in the mm continuum than RW Aur B. The two disks are compact (15-20 au) and misaligned to each other by 12\degr or 57\degr.
    \item The tidal arm discovered by \citet{Cabrit:2006} is part of an apparent circumbinary ring of gas. However, our analysis suggests that the ring-like geometry is a projection effect and is created by the tidal arm around RW Aur A combined with a counter spiral arm around RW Aur B, as predicted in SPH simulations \citep{Dai:2015}. The original tidal stream is at least partially unbound and likely expanding outward from RW Aur A. 
    \item Photometric dimming events have been occurring for $>$80 years with the first event in 1937--1938 and the most recent one still on going. We have identified 7 different dimmings each varying in depth and duration. Long baseline photometry is beginning to explore the same spatial scales as ALMA.
    \item The gas extents of the primary and secondary disks (58 and 38\,AU respectively) are entirely consistent with the most probable values expected to result from tidal truncation. 
    \item RW Aur B has likely undergone multiple eccentric fly-bys of RW Aur A as evident by the multiple tidal streams observed. 
\end{itemize}


Future observations of RW Aur should try understand the influence these star-disk interactions had on the observed high accretion rate of RW Aur A. Additionally, the observed occultation events may suggest that circumstellar material in RW Aur is coalescing to form larger structures. Therefore, it is possible that we are observing the early stages of planet formation. However, it is not clear what potential these small truncated circumstellar disks will have for future planet formation. Higher angular resolution millimeter mapping of the RW Aur A and B disks may provide additional, more conclusive evidence of ongoing planet formation through the presence of gaps in their disks. Our new ALMA observations show the presence of multiple tidal streams of gas which suggest that we are seeing the aftermath of multiple fly-by interactions that have significantly influenced the circumstellar environment in RW Aur. New simulations should try to replicate the observed gas structure and kinematics by allowing for multiple star--disk interactions. Additionally, these simulations should explore the possibility that RW Aur B had an initial disk prior to its first close encounter with RW Aur A.



%



\acknowledgements
Work performed by J.E.R. was supported by the Harvard Future Faculty Leaders Postdoctoral fellowship. RAL gratefully acknowledges funding from NRAO Student Observing Support. T.J.H. is funded by an Imperial College London Junior Research Fellowship. SC and JP acknowledge support from the Programme National ``Physique et Chimie du Milieu Interstellaire'' (PCMI) of CNRS/INSU with INC/INP co-funded by CEA and CNES.

This paper makes use of the following ALMA data: ADS/JAO.ALMA\#2015.1.01506.S and ADS/JAO.ALMA\#2016.1.00877.S. ALMA is a partnership of ESO (representing its member states), NSF (USA) and NINS (Japan), together with NRC (Canada), MOST and ASIAA (Taiwan), and KASI (Republic of Korea), in cooperation with the Republic of Chile. The Joint ALMA Observatory is operated by ESO, AUI/NRAO and NAOJ.

We acknowledge with thanks the variable star observations from the AAVSO International Database contributed by observers worldwide and used in this research. 

\bibliographystyle{apj}

\bibliography{RWAur_ALMA}

\end{document}